\documentclass{aa} 
\usepackage{natbib}
\usepackage{graphicx}
 \usepackage[varg]{txfonts}
\usepackage{natbib}
\usepackage[colorlinks=true,citecolor=blue,linkcolor=blue]{hyperref}

\graphicspath{{./figs/}}
\newcommand{\chandra}{\textsl{Chandra}\xspace}

\begin{document}

\title{Color-color diagrams as tools for assessment of the variable absorption in high mass X-ray binaries}

   \author{
V.~Grinberg\inst{1}\and
M.A.~Nowak\inst{2} \and
N. Hell\inst{3} %\and 
%J.~Wilms\inst{4} \and
%M.~Hirsch\inst{4}
}

\institute{Institut f\"ur Astronomie und Astrophysik, Universität T\"ubingen, Sand 1, 72076 T\"ubingen, Germany \\
  \texttt{grinberg@astro.uni-tuebingen.de}
\and{Physics Department, CB 1105, Washington University, One Brookings Drive, St. Louis, MO 63130-4899, USA}
  \and
  Lawrence Livermore National Laboratory, 7000 East Avenue, Livermore, CA 94550, USA
%  \and 
%Remeis/ECAP
}

\date{ -- / --}

\abstract{High mass X-ray binaries hold the promise of giving us understanding of the structure of the winds of their  supermassive companion stars by using the emission from the compact object as a backlight to evaluate the variable absorption in the structured stellar wind.  The wind along the line of sight can change on timescales as short as minutes and below. However, such short timescales are not available to direct measurement of absorption through X-ray spectroscopy with the current generation of  X-ray telescopes.  In this paper, we demonstrate the usability of color-color diagrams for assessing the variable absorption in  wind accreting high mass X-ray binary systems. We employ partial covering models to describe the spectral shape of high mass X-ray binaries and assess the implication of different absorbers and their variability on the shape of color-color tracks. We show that taking into account the ionization of the absorber, and in  particular accounting for the variation of ionization with absorption depth, is crucial to describe the observed behavior  well.}


   \maketitle

\section{Introduction}

In high mass X-ray binaries (HMXBs) the compact object, whether black hole or neutron star, accretes from a massive 
stellar companion. 
% NH: are they really all wind-fed? I thought some are Roche-lobe overflow? And since you mention Be binaries (do they usually count as HMXB?), aren't those CO smashing through a stellar disk? -- VG: Done
In particular, in Supergiant X-ray binaries (as opposed to Be X-ray binaries), the stellar companion is a 
giant O or B type star whose strong stellar wind feeds the accretion  \citep[for a review, see][]{Martinez-Nunez_2017a}. 
% NH: aren't all HMXB with O/B stars? -- VG: Done
The winds of such stars are line-driven and thus not smooth, but are highly structured, with dense pockets (``clumps'') of gas 
embedded in a rarefied medium \citep[e.g.][]{Owocki_1984a,Feldmeier_1995a}. Observations imply that the clumping sets on already close to the stellar surface, within less than $r \lesssim 0.25\,R_\star$
% NH: can you rephrase that? Are the clumps as close as that to the surface? Or do they start there and stretch out beyond? Or do they only exist within that radius? (r is distance from the surface?) -- VG: Done
\citep{Cohen_2011a,Torrejon_2015a}, i.e., closer than the typical observed locations of the compact objects. Additionally, the presence of the compact object can lead to the formation of large-scale structures in the wind, such as photoionization and accretion wakes or focused wind components \citep{Blondin_1990a}.

Understanding the clumpy structure of O/B stellar winds is crucial both for the understanding of the mass loss, and thus 
evolution, of the O/B stars themselves  \citep{Fullerton_2006a}, and for understanding accretion processes in HMXBs 
\citep{Martinez-Nunez_2017a}. Accretion of individual clumps could lead to the observed flares in both persistent HMXBs 
\citep[e.g.][]{Fuerst_2010a,Martinez-Nunez_2014a,Garcia_2018a} and supergiant fast X-ray transients \citep[SFXTs; e.g,][]{Bozzo_2011a, Ferrigno_2020a}. Clumps passing through 
the line of sight towards the compact object, on the other hand, lead to discrete absorption events 
\citep[e.g.,][]{Balucinska-Church_2000a,Naik_2011a,Yamada_2013a,Hemphill_2014a,Hirsch_2019a}. 
%NH: also cite Hanke 2009? or 2008, but that is the proceedings
%
%The wind clumps will be further deformed during the accretion process close to the compact object \citep{El_Mellah_2018a}.

Models and simulations of the clumpy structure of O/B winds \citep[e.g.,][]{Oskinova_2012a,Sundqvist_2013a,Sundqvist_2018a} 
exist, and so do simulations of clumpy wind accretion onto compact objects \citep{El_Mellah_2018a} and variable absorption in a clumpy wind environment \citep{Grinberg_2015a,El_Mellah_2020a}. However, the 
variability timescales predicted are usually too short to allow current X-ray instruments to accumulate spectra of a quality that would allow to constrain the absorption
during the passage of a single clump well.  
% NH: can we quantify "spectra of good quality" here? That threshold is different for low-res vs high-res. And we haven't talked about which of the two is needed for the characterization of the clumps (guessing high-res really, cause of line diagnostics). Also still would be kinda nice to know what the predicted timescales are here, even though you kinda mention it 2 sentences down. -- VG: changed a bit and tweaked the next sentence with "for example" so that the relationship between the two becomes more clear
For example, \citet{Grinberg_2017a} used the setup of \citet{El_Mellah_2018a} to 
calculate that the correlation time between the peaks of the variable column density for the neutron star HMXB Vela X-1 is at most 
1\,ks, or equivalently only a few self-crossing times of the wind clumps.
% NH: what's a self-crossing time? -- VG: time the thing needs to fly across its own length or the time it needs to pass through the line of sight
A more thorough exploration of parameter space of 
different wind properties in toy models shows that absorption measurements on timescales of a few hundred seconds hold the
potential to allow measurements of clump size and mass \citep{El_Mellah_2020a}. 
% NH: is clump radius a well defined parameter or do we have to talk about what shape we assume first? -- VG: changed to "size"; we assume spherical clumps on the above, everything else needs to be separately simulated or very cleverly addresses analytically; something for the next project
While lightcurves and hardness ratios during 
dipping show variability on such  timescales and below \citep{Grinberg_2017a,Hirsch_2019a}, the periods are too short to 
accumulate well defined spectra and directly measure the absorption in a single dip event of a few 100\,s length or below from current instruments' spectra.

%XXX Maria's \& Ivi's papers, dips visible by eye with lengths that can be as short as below a minute . XXX 

%NH: Hanke once had a Bachelor student (Strobel) check the dip lengths in RXTE data. There is short mention in Hanke 2011, but other than that just the Bachelor thesis, I think. 

%This is too short a timescale to accumulate spectra with current instruments. May accessible with Athena but not clear. Emphasize why we need to disentangle different absorption levels:  washing out of lines etc., example from Vela, also mention Cyg X-1.

However, where not enough information is available for a full spectral analysis, location in (X-ray) color-color space can be 
used to obtain information about spectral properties. Similar approaches have been employed when analyzing faint sources and, 
e.g., when trying  to understand source populations in other galaxies or contributions from different emission components 
\citep[e.g.,][]{Carpano_2005a}. The advantage of HMXBs is that the underlying spectral shape of the continuum emission from 
the compact  object can often be well constrained from time-averaged spectra, especially when data at high energies not  
% NH: define high energy. >4keV and <10keV? Anything below 20 keV? probably not as much as 100 keV. 
affected by absorption ($\gtrsim 10$\,keV) are available, allowing for better constraints on the expected location in color-color space.

The aim of this paper is thus to connect the shape of tracks on color-color tracks with absorption variability in HXMBs.
%The aim of this paper is thus to discuss the usability of tracks in X-ray color-color diagrams to constrain absorption variability in HMXBs.
It builds upon previous work of \citet{Nowak_2011a} and \citet{Hanke_2011_PhD} who attempted to describe the color-color tracks of neutral absorbers and expands it with more realistic, ionized absorbers 
that are a better description for the hot, structured winds of O/B stars.

%Working towards an observational method to assess absorption variability on timescales that are not accessible with X-ray spectroscopy is crucial for 
% NH: ehm, do we know how much time we need to populate a CC-diagram with enough statistics that this is useful? I am guessing shorter time scales than trying high-res line diagnostics, but probably still something like, I dunno, 100 lightcurve bins? At 25s (e.g., chandra. Or did we do 25ms? No, I think it was seconds) that is still like 2.5 ks >> 1 ks. 
%comparisons of the observed absorption variability to that predicted from state of the art simulations of clumpy wind accretion. \textbf{ADRESS NATALIE'S CONCERNS}
%\citep[e.g.,][]{El_Mellah_2018a}. In particular, EL MELLAH IN PREP develop the theory that would allow to use short-term variability of the absorbing column to measure clump masses and radii, but the timescales in question (100s of seconds and below) cannot be accessed through direct spectral fits.
We first discuss the observational signatures of dipping in wind-accreting HMXBs and partial covering models often used to described HMXB 
spectra in Sec.~\ref{sec:signatures}. We then re-visit variable neutral absorber models in Sec.~\ref{sec:neutral}, where we additionally 
discuss the influence of covering fraction and changes in underlying spectral shape on the shape of color-color tracks. In 
Sec.~\ref{sec:warmabs}, we discuss the effects of a warm absorber and in particular the effects of a 
warm absorber whose ionization depends on its equivalent hydrogen column density, as would be expected in a wind environment with clumps 
or other overdensities. We finish with a summary and outlook in Sec.~\ref{sec:outlook}.

\section{Signatures of Dipping in wind-accreting HMXBs}
\label{sec:signatures}

\subsection{Observational patterns}

Short, pronounced increases of absorption, often called dips, have been observed in a number of wind-accreting HMXB sources such as Cyg~X-1 \citep[e.g.][]{Li_1974a,Balucinska-Church_2000a,Hirsch_2019a}, Vela~X-1 \citep{Odaka_2013a,Grinberg_2017a}, and 4U 1538$-$52 
\citep{Hemphill_2014a}.
% NH: so you are saying, I need to go extract the lightcurves of those sources in the Chandra proposal...? Also, what about Felix' OAO source? That seems pretty variable.
% VG - you may also just remember the literature better than I do, I was rather hoping for this ;)
%
% potentially:
% Cen X-3: http://tgcat.mit.edu/tgPlot.php?t=P&i=3690  (but has disk winds and a warped disk; other two obs are eclipse and egress)
% 4U 1908_075: http://tgcat.mit.edu/tgPlot.php?t=P&i=3968
%              http://tgcat.mit.edu/tgPlot.php?t=P&i=3931
% 4U 1907+09: http://tgcat.mit.edu/tgPlot.php?t=P&i=5588
%              http://tgcat.mit.edu/tgPlot.php?t=P&i=5587
% 4U 1700-37: http://tgcat.mit.edu/tgPlot.php?t=P&i=5388 (flares?)
%
% a little variable, but wouldn't call it a dip:
% OAO 1657-415: http://tgcat.mit.edu/tgPlot.php?t=P&i=4454
% 
As opposed to off states, which are interpreted as cessation of accretion, the underlying continuum does not change during dips and the 
spectral variability is driven by changes in the absorbing column. The length of such absorption episodes ranges from 10s of seconds to ks and above, with the 
longer dipping episodes often showing pronounced temporal sub-structure.

%\begin{itemize}
%\item 3x example plots from different sources \& instruments (Chandra,
%  XMM, Suzaku?)
%\end{itemize}

So far, color-color diagrams have been mainly used to disentangle the absorption level in the bright HMXB Cyg~X-1 with a variety of 
different X-ray instruments \citep{Nowak_2011a,Miskovicova_2016a,Basak_2017a,Hirsch_2019a}. It is the most prominent example as it is 
bright and well studied as a key source for understanding both stellar winds in HMXBs and the physics of accretion onto black holes. Both 
\citet{Nowak_2011a} and \citet{Hanke_2008a} attempted, with some success, a physical description of the color-color tracks based on \textsl{Suzaku} and 
\textsl{Chandra} observations, respectively,  with a partial covering model (see Sec.~\ref{sec:partialcoverer}).
However, discrepancies between the observations and data remain; in particular, the curve of the data is more pointy than that of 
the model. \citet{Nowak_2011a} attributed this to possible influence of ionization, but did not attempt to include an ionized absorber 
into their  models. From high resolution spectra, the presence of an ionized absorber has been shown by \citet{Hanke_2009a} and 
\citet{Miskovicova_2016a}  and its variability with increasing level of absorption in \citet{Hirsch_2019a}. Note that \citet{Hirsch_2019a} 
used color-color diagrams  to divide their observation into different absorption levels; however, they described the shape of the tracks 
with  an empirical polynomial model only.

\subsection{Partial covering model} \label{sec:partialcoverer}

The simplest assumption for a HMXB spectrum is a continuum modified by
a partially covering absorber, i.e., only a certain percentage (the
covering fraction, $f_\mathrm{c}$) of the X-rays emitted by the central
source are absorbed locally while the rest ($1-f_\mathrm{c}$) arrives
at the observer modified by interstellar absorption only. Such models 
have been used widely in the literature to describe HMXBs 
\citep[e.g.,][]{Fuerst_2014a,Fornasini_2017a}. The partial
covering can be realized in different ways:

\begin{itemize}

\item partial covering in space: the absorber covers only a part of
  the emitting region. This could be the case for a relatively small, compact
  absorber and/or an emission region that is extended when compared to
  the absorber.
  % NH: what's considered small is distance dependent a bit. See solar eclipse

\item partial covering in time: any observation averages over some
  exposure time. If the absorber changes during this time (as would be
  expected given the quick dynamic timescales of both structured
  stellar wind and accretion streams), the observed emission is the
  sum of spectra with different levels of absorption.

\item dust scattering \citep[e.g.][and references therein]{Xu_1986a}:
  the spectrum consists of a directly observed component that is
  subject to local absorption events and a somewhat time-delayed and
  averaged (due to scatterings at different radii) spectrum from the
  dust scattering halo. This spectrum will be softer than
  the direct emission, due to the energy-dependency of the scattered
  emission. 
  %NH: didnt Hanke talk about dust scattering somewhere? Or was it Nowak?

\end{itemize}

The situation may be further complicated through the existence in some
sources of a soft excess component of still often unclear origin
\citep[e.g.,][]{Hickox_2004a} that may be affected by absorption
differently than the primary continuum components. This soft excess includes possible contribution from a photoionized wind component, where a forest of unresolved fluorescence lines could contribute to the overall soft emission on scales much larger that the point-like emission from the vicinity of the compact object.

A basic model to describe partial covering can be written as:\begin{equation}
\texttt{abs}_\texttt{ism} \times \texttt{continuum} \times (f \times
\texttt{abs}_\texttt{wind} + (1-f)) \label{eq:pc}
\end{equation} with $\texttt{abs}_\texttt{ism}$ the absorption in the interstellar 
medium (ISM), \texttt{continuum} the continuum emission, $0 \leqq f \leqq 1$ the 
covering fraction and $\texttt{abs}_\texttt{wind}$ the local absorption in the 
system, i.e., in the (disrupted) stellar wind of the companion. For the continuum 
at energies $\lesssim$10\,keV, a power law usually offers a good description. 
In general, physical models for the X-ray continuum emission in HMXBs, usually from highly 
magnetized neutron stars, are rare and the continuum is described empirically 
by power laws, modified by different breaks and cutoffs, or Comptonization models 
that are power-law shaped. Similar empirical descriptions can be employed for 
black hole accretors.

\begin{figure}
\includegraphics[width=\columnwidth]{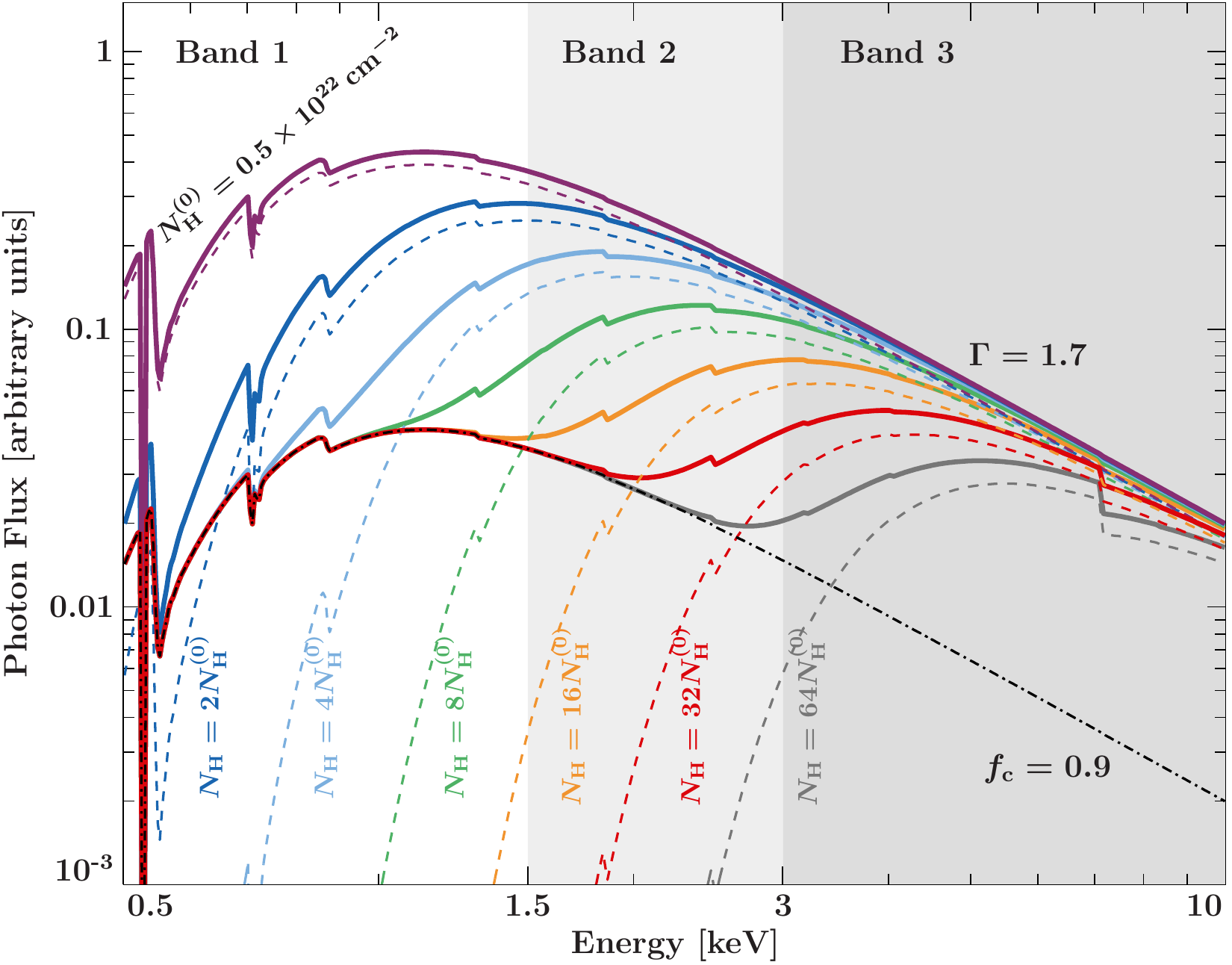}
\caption{
In principle demonstration of the effect of a variable partial 
covering absorption on an absorbed powerlaw with photon index $\Gamma=1.7$ and ISM absorption $N_\mathrm{H}^{0}$. We assume a covering fraction of 0.9, and a varying  equivalent hydrogen column density, $N_\mathrm{H}$, of the partial coverer, ranging from $N_\mathrm{H}^{0}$, implying only ISM absorption, to $128 N_\mathrm{H}^{0}$. The uncovered fraction (here, $1- f _\mathrm{c} = 0.1$), is shown as a black dash-dotted line. The covered, absorbed component is shown in dashed lines, with colors corresponding to different $N_\mathrm{H}$. The sum of the covered and uncovered fractions is shown in the solid colored lines.
With increasing $N_\mathrm{H}$ of the partial coverer, we first see diminution primarily of Band 1. 
The partial covering fraction leads to a``floor'' for the flux value in band 1, while band 2 and band 3 continue to decrease with increasing column.
}
\label{fig:abc}
\end{figure}

Such a model for a partial covering can be easily set up in most current X-ray 
analysis environments; in our work we use the Interactive Spectral Interpretation 
System (ISIS) version 1.6.2 \citep{Houck_Denicola_2000a,Houck_2002,Noble_Nowak_2008a}. 
Here and in the remainder of the paper, we discuss the effects of a partial coverer 
using a power law continuum -- such a simple continuum description has been used 
repeatedly in the literature to describe the continuum spectral shape of HMXBs below 10\,keV 
\citep[e.g.,][]{Hemphill_2014a,Miskovicova_2016a,Grinberg_2017a}. Realizations of 
Eq.~\ref{eq:pc} for a covering fraction $f = 90\%$ and different values of local 
absorption are shown in Fig.~\ref{fig:abc}, which follows a similar illustration in \citet{Hirsch_2019a}.

For color-color diagrams, we consider the fractions (``colors'') between the 
fluxes in three energy bands, with band 1 having the lowest and band 3 the highest 
energies with soft color 1/2 and hard color 2/3. It can easily be seen in Fig.~\ref{fig:abc} 
that the value of a given color will not be a monotonic function of absorption strength. 
In particular, the soft color 1/2 will be similar when the local absorber $\texttt{abs}_\texttt{wind}$ is absent and when the local, 
partially covering absorption is very strong and such a strong absorber will remove all covered flux in the bands 1 and 2, leaving just the contribution of the uncovered component.

\section{A variable neutral absorber}
\label{sec:neutral}

Here, we introduce the tracks that a source traces in the color-color diagram in the 
presence of a variable neutral absorber. Some effects of the variable neutral absorber on such tracks have been discussed in the previous works by \citep{Hanke_2008a}, \citet{Nowak_2011a}, and \citet{Hirsch_2019a} 
and we extend on these first discussions by including the influence of short-term 
variability of spectral shape of the underlying continuum.

\subsection{Tracks on color-color diagrams}
\label{sec:neutraltracks}

\begin{figure*}
\includegraphics[width=\textwidth]{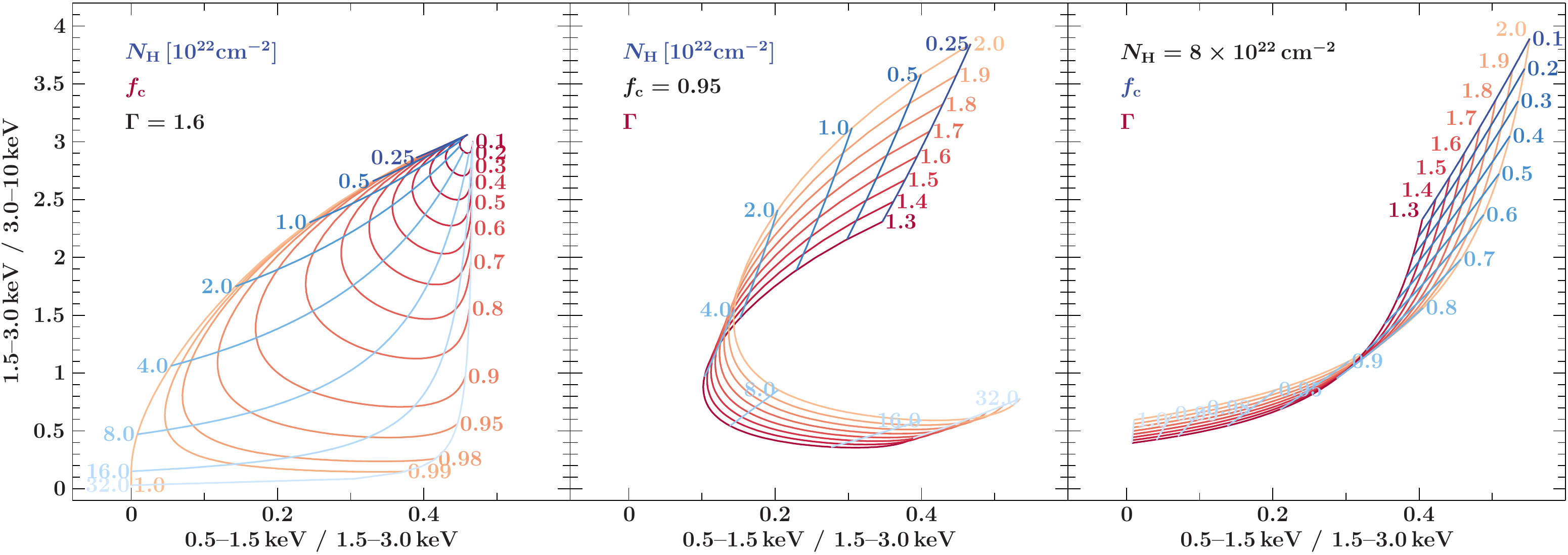}
\caption{Simulated color-color tracks for varying parameters of a partially absorbed power law model for HETG-MEG and parameter ranges typical for the HMXB Cygnus X-1.}\label{fig:neutralgrid}
\end{figure*}

In the partial covering model presented in Sec.~\ref{sec:partialcoverer}, there are three main parameters that define the shape of
the spectrum and can vary: the photon index $\Gamma$ of the power law continuum,
the absorbing column of the local absorber
$\texttt{abs}_\texttt{wind}$ and the covering fraction $f$. The absorption in the interstellar medium, $\texttt{abs}_\texttt{ism}$, is constant.

The photon index can show intrinsic variability on both short and long scales \citep[e.g.][]{Skipper_2013a,Grinberg_2013a,Fuerst_2014a}, although we point out that strong changes are usually associated with changes in emission geometry and thus happen on longer timescales than those addressed in this work. In the clumpy wind or clumpy absorber paradigm the local absorber can change strongly on short timescales below minutes, as clumps enter and leave the line of sight towards the compact object \citep[e.g.,][]{El_Mellah_2020a}. Short-term changes in the covering fraction are somewhat harder to realize, but could be due to changing number of clumps along the line of sight or the time delay between changes in the primary continuum and scattered component. 

The exact shape of the tracks is always going to be dependent on a given instrument, even for the same spectral shape of the source. Here and the following, when discussing general trends (Sec.~\ref{sec:homogentracks} and~\ref{sec:inhomogentracks}), we calculate the tracks that such changes lead to in a color-color diagram using the example of a \textsl{Chandra}-HETGS/MEG observation 
% NH: didn't you specifically use one order of MEG? maybe should admit that
of a bright source using typical value ranges for $\Gamma$ and $f_\mathrm{c}$. We model the absorption with the updated version of the \texttt{tbabs}\footnote{This model was formerly known as \texttt{tbnew} and is since 2016 included as the default tbabs version in \texttt{xspec} \citet{Arnaud_1996a} and thus also all packages that rely on xspec models such as ISIS.} model \citep{Wilms_2000a}, using \texttt{wilm} \citep{Wilms_2000a} cross-sections and \texttt{vern} abundances \citep{Verner_1996a}. We simulate observed spectra using real \textsl{Chandra}-HETGS/MEG responses (RMFs and ARFS), use the so simulated observations to calculate hardness values, and show the resulting tracks in Fig.~\ref{fig:neutralgrid}. 

\begin{figure*}
    \centering
     \includegraphics[width=0.31\textwidth]{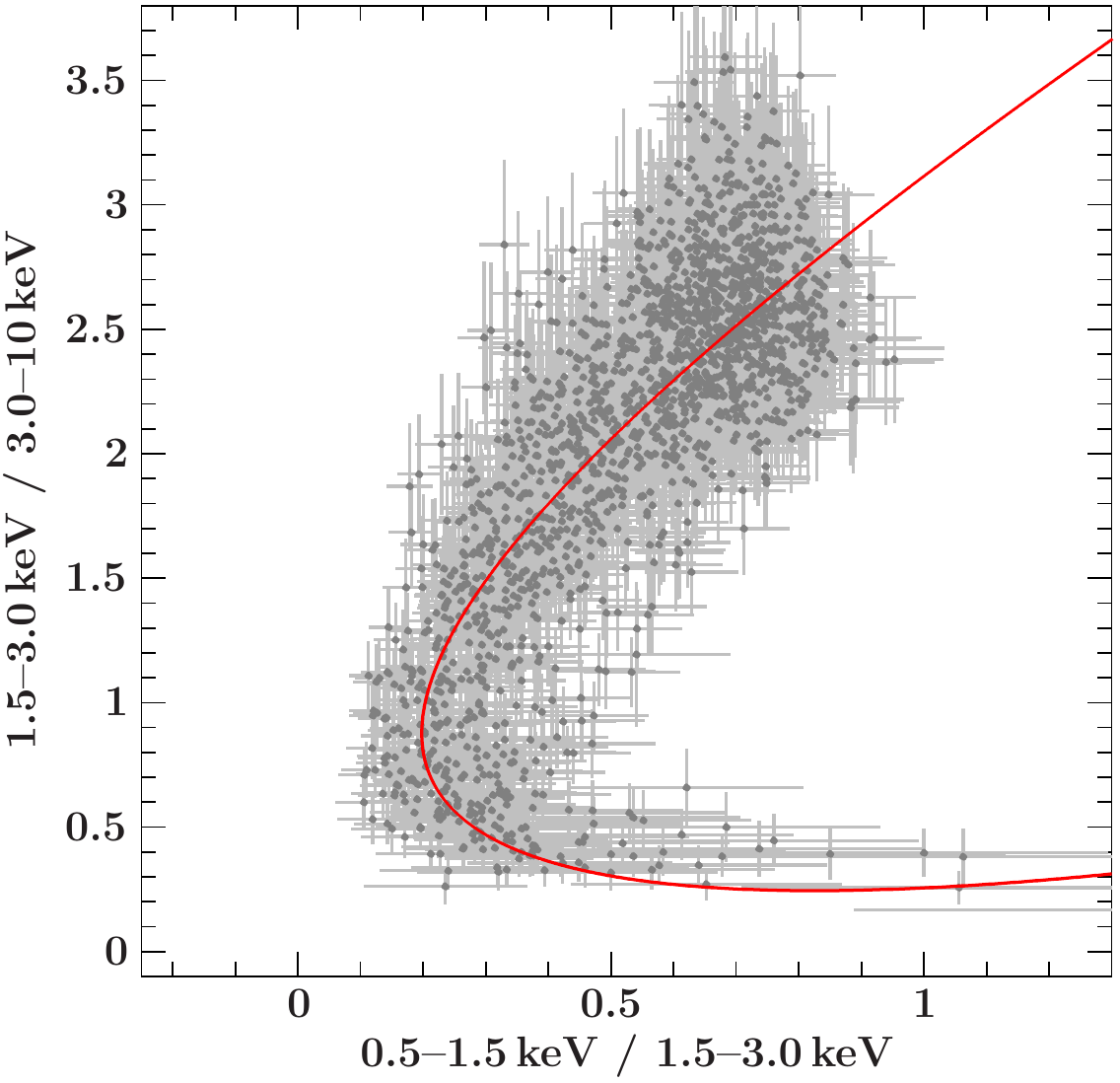}\hfill
    \includegraphics[width=0.31\textwidth]{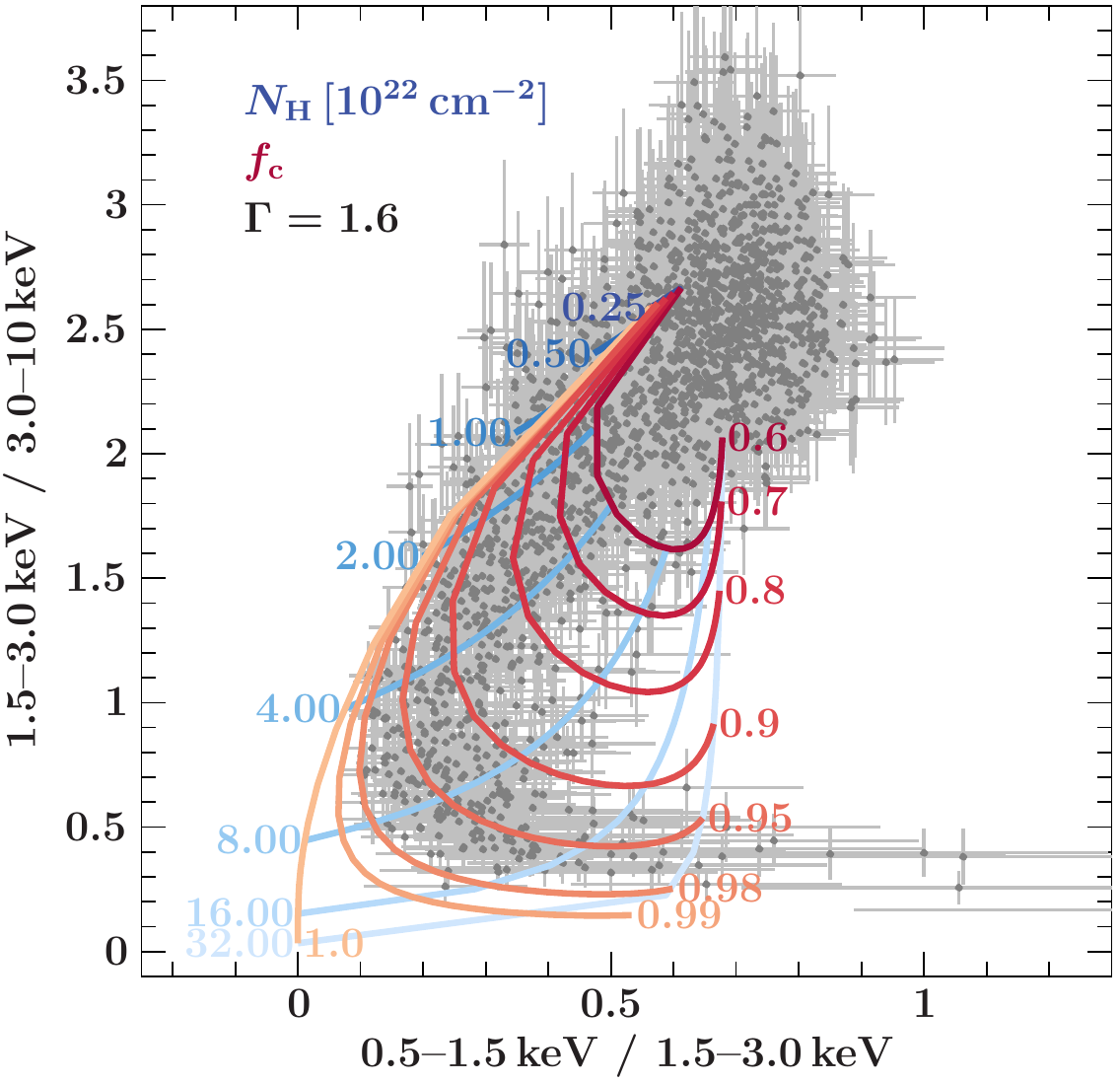}\hfill
        \includegraphics[width=0.31\textwidth]{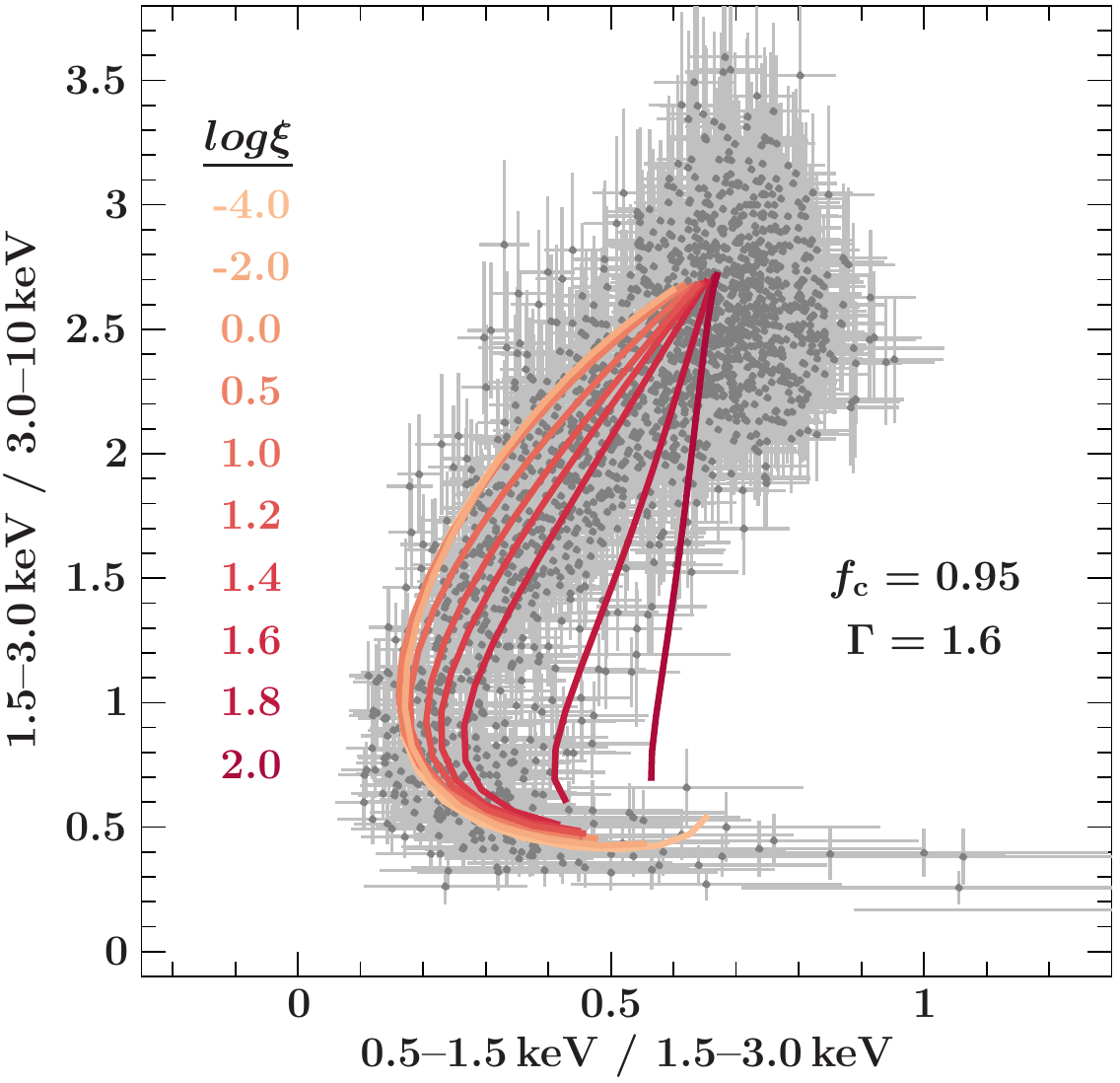}
    \caption{Comparison of the color-color diagrams of the \chandra observation ObsID~3814 using the MEG M1 arm with different models. \textsl{Left:} Empirical polynomial fit. \textsl{Middle:} Neutral grid using the \texttt{tbabs} absorption model for different covering fractions and the same continuum photon index $\Gamma = 1.6$. \textsl{Right:} Homogeneously ionized absorber using the \texttt{warmabs} model with continuum photon index $\Gamma = 1.6$ and a covering fraction of 0.95. $N_\mathrm{H}$ is changing from $0.25 \times10^{-22}$\,cm$^{-2}$ to $32\times10^{-22}$\,cm$^{-2}$ along each line, with each line showing the track for different ionization parameter $\log \xi$ as indicated by the different colors.}
    \label{fig:chandra}
\end{figure*}

Changes in individual parameters lead to typical changes in the color-color tracks: changes in the neutral absorption column density lead to typical curved or ``nose-shaped'' tracks (Fig.~\ref{fig:neutralgrid}, left and middle), while changes in covering fraction and intrinsic spectral shape show distinctly different patterns that could lead to some of the spread around the curves that observational data show (Fig.~\ref{fig:chandra}). If the underlying shape of the color-color track can be well modelled, such deviations can be potentially used to assess the variability of the underlying continuum and covering fraction on short timescales. As the driving observable is the actual shape of the nose-shaped track, we will concentrate on changes in absorption column for the remainder of the paper.

\subsection{Comparison to observed color-color tracks for Cyg~X-1}
\label{sec:neutcyg}

A variable neutral absorber is a simple enough model that can be used to directly model observed color color diagrams. \citet{Nowak_2011a} have presented fits of such a model to \textsl{Suzaku} data and \textsl{Chandra}-HETGS data have been discussed by \citet{Hanke_2008a} and \citet{Hanke_2011_PhD}. In all cases, the model describes the overall trend of the shape of the tracks, but fails to reproduce the details; in particular, the curvature of the data is more pointy than the model. 

For a comparison with the data, we use \textsl{Chandra} ObsID~3814, a $\sim$50\,ks observation taken in the TE graded mode on 2003-04-19 and covering the orbital phases 0.93-0.03, i.e., at a phase where strong dipping is expected and indeed observed. To obtain lightcurves with a resolution of 25.5\,s, we follow the standard extraction procedure, as, e.g., also done in \citet{Hirsch_2019a}. The high resolution spectrum of the non-dip phase of this observation has been previously analyzed by \citet{Hanke_2009a}, whose analysis also includes the simultaneous broadband data taken with the \textsl{R}XTE satellite. \citet{Hanke_2009a} list different values in the range of 1.5--1.7 for the photon index of the power law used to describe the continuum and show the importance of coverage above $\sim$10\,keV to constrain the continuum shape, but also emphasize the remaining uncertainties in spectral modelling. The high-resolution spectra of the Si and S regions of the observation have been analyzed by \citet{Hirsch_2019a} during dip- and nondip-phases and for our theoretical predictions, we use the same MEG minus 1 order RMF and ARF files for non-dip phases as they did in their analysis.

To guide the eye, we first characterize the shape of the tracks using an empirical fit, in an approach similar to \citet{Hirsch_2019a} (Fig.~\ref{fig:chandra}, left panel). Specifically, we characterize the curve using a parameterized second-degree polynomial for each of the two colors. The best fit is obtained by minimizing the distance of each data point to the curve. The distances in either color are weighted with the respective uncertainty of each data point as in the definition of $\chi^2$ statistics. 
%\textbf{Vici to check phrasing. More details in comment for reference.}
%-- VG: I think this is good! thanks!
% quote from Hirsch+2019:
% "We therefore determine the shape of the track empirically by fitting an empirical curve to the scatter plot. This curve is described through a parameterized polynomial of second degree for each of the two colors in the diagram. To find the polynomial coefficients, a χ 2 minimization algorithm is used to optimize the shortest distance of each data point to the curve."
% Additional info:
% we minimize:  f[i] = min( ((ab[i]-xt)/ab_err[i])^2 + ((bc[i]-yt)/bc_err[i])^2 );
% where (ab[i], bc[i]) are each point in the CC diagram, and (xt, yt) are arrays with the x and y values for the current iteration of the curve.
% f is the return value of the fit function and fitted against zeros. 

We then turn to comparing the data with theoretical color-color tracks for a neutral absorber. The tracks for $\Gamma = 1.6$ are shown on the middle panel of Fig.~\ref{fig:chandra}. We also calculate the tracks for $\Gamma = 1.5$ and 1.7 and compared them to the observations. This 
% NH: I didn't not understand why this comment is here? Did you do anything with 1.5 and 1.7? It's not in the figure (would make it illegible). But not sure what I am supposed to glean from this info 
%VG: explained this better now - hopefully
resulted in the expected slight shift of the tracks mainly along the vertical axis ( as expected from Fig.\ref{fig:neutralgrid}, middle panel), with $\Gamma = 1.6$ agreeing with the data best. The behavior we observe is consistent with the problems previously pointed out in the literature: the data is more pointy than the predictions.
 %NH:  one of the Cyg X-1 observationas (8525, I think) also has that mostly unpublished multi-satellite campaign
%VG: rue, but JW is gonna kill be if I touch this one + I don't actually want to analyze everything myself here
This deviation cannot be explained by variability of power law shape or covering fraction (cf. Fig.~\ref{fig:neutralgrid}). \citet{Nowak_2011a} suggested that the deviation stems from the assumption of a neutral absorber: the wind material is intrinsically ionized to some level \citep[e.g.,][]{Sander_2018a} and any material in the vicinity of the compact object will be further ionized by its intense X-ray radiation. We discuss such an ionized absorber in the next section.

\section{Warm, ionized absorbers} \label{sec:warmabs}

\subsection{Modelling an ionized absorber}

The ionization of the material local to the HMXB changes its X-ray absorbing properties. In the following, we will assess how the change in ionization influences the tracks a source traces on the color-color diagram with variable absorption.

\begin{figure}
\includegraphics[width=\columnwidth]{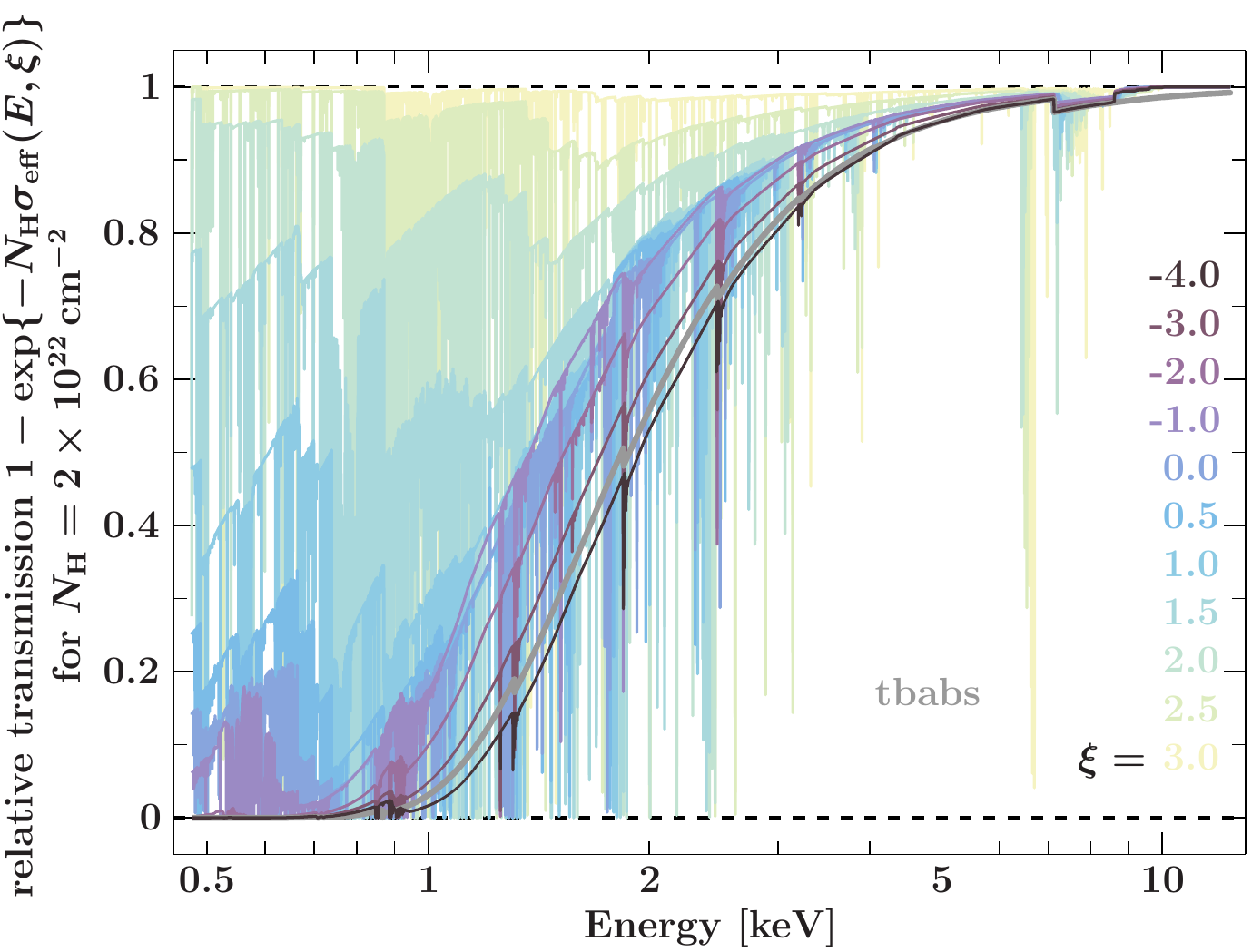}
\caption{
% NH: cream might not quite be the right color for a paper
Relative transmission in the 0.5--10\,keV band for an equivalent hydrogen column density of $N_\mathrm{H} = 2\times10^{22}\,\mathrm{cm}^{-2}$ and varying ionization parameters using \texttt{warmabs} shown in color. For comparison, neutral absorption as modelled with the \texttt{tbabs} model for the same hydrogen column density in shown in gray.}\label{fig:warmabs_specs}
\end{figure}

To do so, we are using version 2.30 of the \texttt{warmabs} family of photoionization models \citep{Kallman_2009a}, a part of the \texttt{XSTAR} package \citep{Bautista_2001a,Kallman_2001a}, to model the ionized absorber.
Fig.~\ref{fig:warmabs_specs} shows how the relative absorption as calculated by \texttt{warmabs} changes for the same equivalent hydrogen column density with varying ionization parameter, $\log \xi$, with
\begin{equation}
  \xi = L_\mathrm{x} / n r^2 \label{eq:xi}
\end{equation}
 defined after \citet{Tarter_1969a} with $L_\mathrm{x}$ the ionizing luminosity above 13.6\,eV, $n$ the gas density and $r$ the distance from the ionizing source. The \texttt{warmabs} model can be used to access ionization parameters $\log \xi$ between -4.0 and 5.0.  While the relative absorption decreases overall, the decrease in individual bands is a complex function of ionization due to contributions of different elements and ions (Fig.~\ref{fig:warmabs_specs}). We note that for the same column density, the lowest ionization of \texttt{warmabs} ($\log \xi = -4.0$) results in a stronger absorption than the neutral \texttt{tbabs} with abundances and cross-sections as discussed in Sec.~\ref{sec:neutraltracks}. This difference is due to the different abundances and cross-sections used in the models. 
 %and can be disregarded for the broad-band approach employed in this work.
 % VG: Manfred discussed the differences in his thesis, but I really don't think that we need to discuss them here
 
To account for standard conditions and for ease of reproducibility, we are using standard population files delivered with \texttt{warmabs}. In particular, we use the files pre-calculated for densities of $10^{12}\,\mathrm{cm}^{-3}$, 
% NH: that's electron densities, right? -- VG: no, it's number density
as these are typical densities of the smooth wind at the location of the compact object \citep{Lomaeva_2020a} and as \texttt{warmabs} using these population files has been previously successfully used to model the observations of Cyg X-1 during the non-dipping phase \citep{Hanke_2011_PhD}. Though we note that densities in a structured wind can show a gradient up to a factor 1000 and above \citep[e.g.,][]{Sundqvist_2018a}. Standard populations files are calculated for an illumination with a powerlaw with a photon index of $\Gamma = 2$. While this is softer than the $\Gamma = 1.6$ powerlaw we employ here, the differences can be neglected given the broad energy bands used and further uncertainties, such as the above mentioned density gradients and the likely presence of a multi-phase medium in wind-accreting HMXBs \citep[e.g.,][]{Boroson_2003a,Grinberg_2017a, Lomaeva_2020a}. 

\subsection{Homogeneous warm absorber}

\subsubsection{ Tracks on color-color diagrams}
\label{sec:homogentracks}

\begin{figure*}
\includegraphics[width=\textwidth]{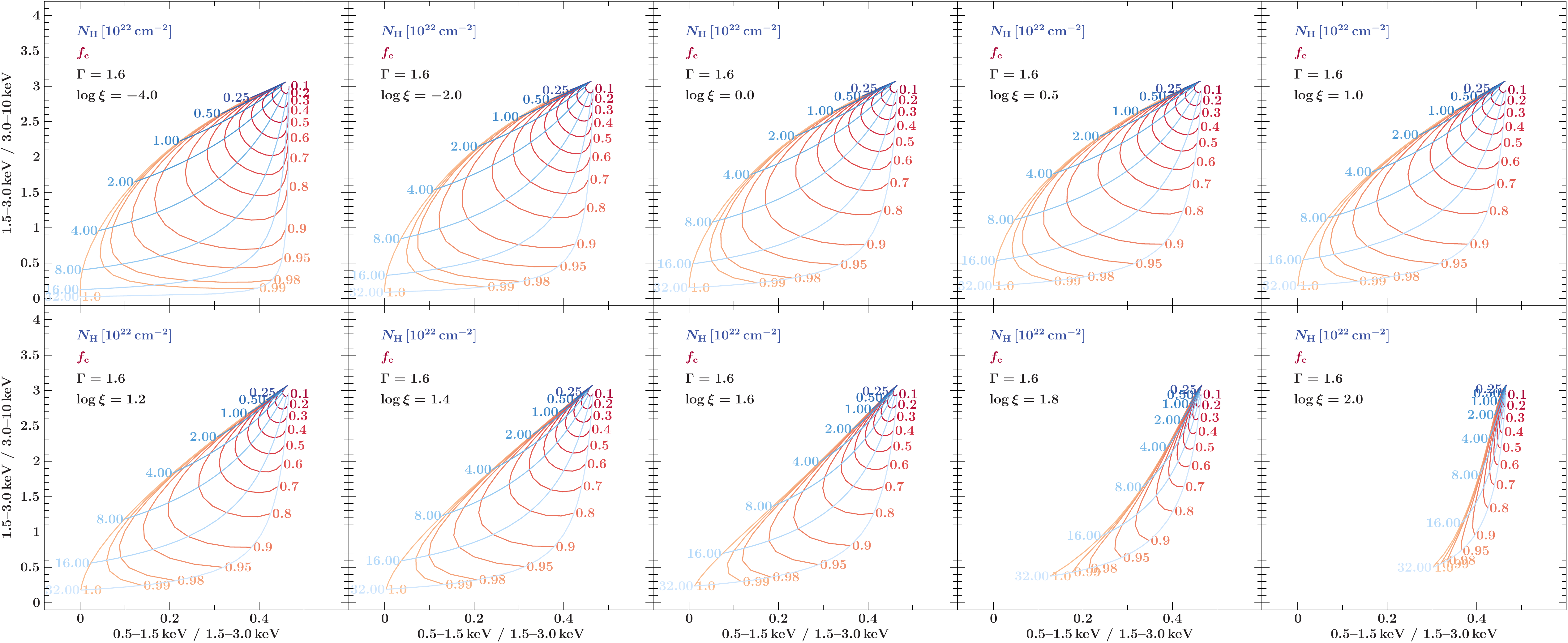}
\caption{Homogeneous warm absorber grids for HETG-MEG with typical Cyg X-1 continuum parameters for different ionization parameters $\log \xi$. Each grid is calculated for varying equivalent hydrogen column density and covering fraction.}\label{fig:warmgrid}
\end{figure*}

We calculate the tracks on the color-color diagrams for media of different ionization. Different ionization levels of the warm absorber lead to changes in the shape of the tracks in the color-color diagrams, as shown in Fig.~\ref{fig:warmgrid}.  For the lowest ionization parameter, $\log \xi = -4.0$, the tracks are similar to those of a neutral absorber modelled with \texttt{tbabs}, as expected.
As transmission at lowest energies changes strongly with increasing ionization parameter (cf. Fig.~\ref{fig:warmabs_specs}), the 0.5--1.5\,keV/1.5--3.0\,keV ratio is most affected. This leads to less pronounced curvature in the tracks.
For the X-ray colors used here, namely 0.5--1.5\,keV/1.5--3.0\,keV and  1.5--3.0\,keV/3.0--10\,keV ratios, the changes in the shape of the tracks with varying $\log \xi$ are strongest for ionization parameters in the range of $\log \xi \approx 1.0$--$2.0$. 

These values are to be compared to those expected from observation -- usually high resolution spectroscopy analyses -- of hot plasmas in HMXBs.
In the  analysis of Cyg X-1 at different absorption stages by \citet{Hirsch_2019a}, the observed intermediate and lower ionization states of silicon and sulfur point towards ionization parameters in the order of 1--2, i.e., the colder denser regions in the wind have values of $\log \xi$ comparable to the ones presented in Fig.~\ref{fig:warmgrid}. Similarly, \citet{Lomaeva_2020a} model the multicomponent plasma in Vela X-1 with a model consistent of two photoionization \texttt{cloudy} models, the colder of which has $\log \xi \approx 1.7$. In 4U~1700$-$37, \citet{Haberl_1989a} found $\log \xi \approx 1.6$, somewhat variable between flaring and off-flaring phases. For the same source, observation with \textsl{Chandra}-HETG analyzed in \citet{Boroson_2003a} imply a range of possible ionization parameters, depending on whether a pure photoionization ($\log \xi \approx$\,2.5--3) or hybrid plasma ($\log \xi \approx$\,1.6) is assumed, but also imply simultaneous presence of lower ionization plasma with $\log \xi \lesssim 1$ in the system.
 
\subsubsection{Comparison to observed color-color tracks for Cyg X-1}

We show comparisons of the warm absorber tracks with \chandra data in the third panel of Figure~\ref{fig:chandra}, using the same observation as in Sec.~\ref{sec:neutcyg}. We calculate the grids for $\Gamma = 1.6$ and choose a covering fraction of $f_\mathrm{c} = 0.95$, based on the tracks that describes the data well in Sec.~\ref{sec:neutcyg}. A mildly ionized absorber ($\log \xi \approx 1.4$) describes the data better than a neutral or quasi-neutral absorber, in agreement with previous results discussing high resolution spectra of the same observation \citep{Hanke_2009a, Hirsch_2019a}. Overall, however, the ionized tracks appear rather flatter than more pointy and therefore do not solve the general problem presented when trying to model color-color diagrams \citep{Nowak_2011a}.

\subsection{Warm absorber with an ionization gradient: structured clumps}
\label{sec:varxi}

In our modelling so far, we have made the assumption that the ionization of the local absorber is constant. 
A realistic structured stellar wind -- no matter whether the structure is due to intrinsic wind clumping or due to the presence of a clumpy accretion wake or similar structures -- will not have a constant ionization.  Confronted with the same ionizing flux, a thicker or denser clump, will be less ionized (Eq.~\ref{eq:xi}).

Such variable ionization has been detected in Cyg X-1 by \citet{Hirsch_2019a}, who conclude that the clumps show ionization structure, with less ionized cores and more ionized outer parts, covering ionization parameters in the range of $\log \xi \approx$~1--2. \citet{Hirsch_2019a}, also point out that simple photoionization is likely not enough to explain the observed ionization structure.
%, however a detailed discussion of the origin of the ionization structure, e.g., in the shocks in the wind, is out of scope of both theirs and this work.

In color-color diagrams, comparatively small changes in $\log \xi$ can lead to strong changes in the shape of the tracks. For the instrument and energy ranges chosen here, this is especially the case for ionization parameters in the range of $\log \xi \approx$~1--2 (Fig.~\ref{fig:warmgrid}). 
We will thus relax the assumption of a constant ionization parameter in the following and investigate track shapes for a variable ionization.

\subsubsection{Tracks on color-color diagrams}
\label{sec:inhomogentracks}

\begin{figure}
    \centering
    \includegraphics[width=0.7\columnwidth]{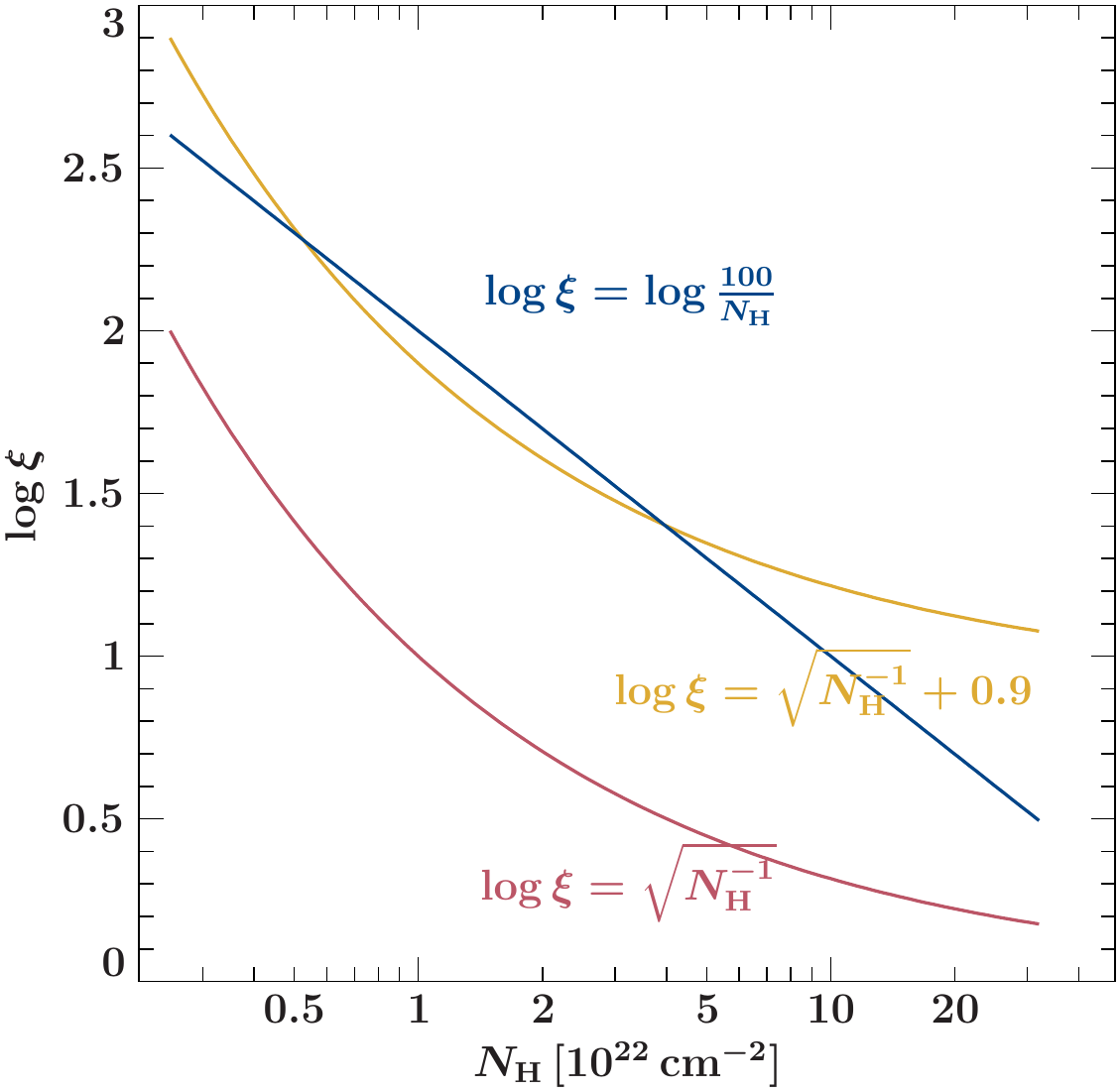}
    \caption{Examples of possible dependencies of the ionization parameter, $\log \xi$, on the equivalent hydrogen absorption column density, $N_\mathrm{H}$, expressed in units of $10^{22}\,\mathrm{cm}^{-2}$.}
    \label{fig:logxiofnh}
\end{figure}

While the general trend of lower ionization of the absorbing material with increasing absorption column is clear, the exact dependency of $\log \xi$ on $N_\mathrm{H}$ is not known. We thus discuss three possible cases as shown in Fig.~\ref{fig:logxiofnh}. First, building on the definition of of ionization parameter (Eq.~\ref{eq:xi}) we assume a function of the form $\log \xi = \log (C_1 / N_\mathrm{H})$ with $C_1$ a constant that we choose equal to 100 and
% NH: is there any significance to why 100? 
$N_\mathrm{H}$ in units of $10^{22}\,\mathrm{cm}^{-2}$ (case 1).  The constant $C_1$ is chosen so that  $\log \xi$ covers the range between 2.5 and 0.5, for the interesting range of $N_\mathrm{H} = 0.25$--$32\times10^{22}$\,cm$^{-2}$. Our second approach is a function of the form $\log \xi = N_\mathrm{H}^{-0.5} + C_2$, assuming $C_2$ as 0 (case 2) and 0.9 (case 3), again with $N_\mathrm{H}$ in units of $10^{22}\,\mathrm{cm}^{-2}$. In cases 2 and 3 the ionization parameter remains higher even at high column density of the absorber, as would be the case in the presence of an additional ionization process, e.g., collisional ionization.

\begin{figure*}
    \centering
     \includegraphics[width=0.31\textwidth]{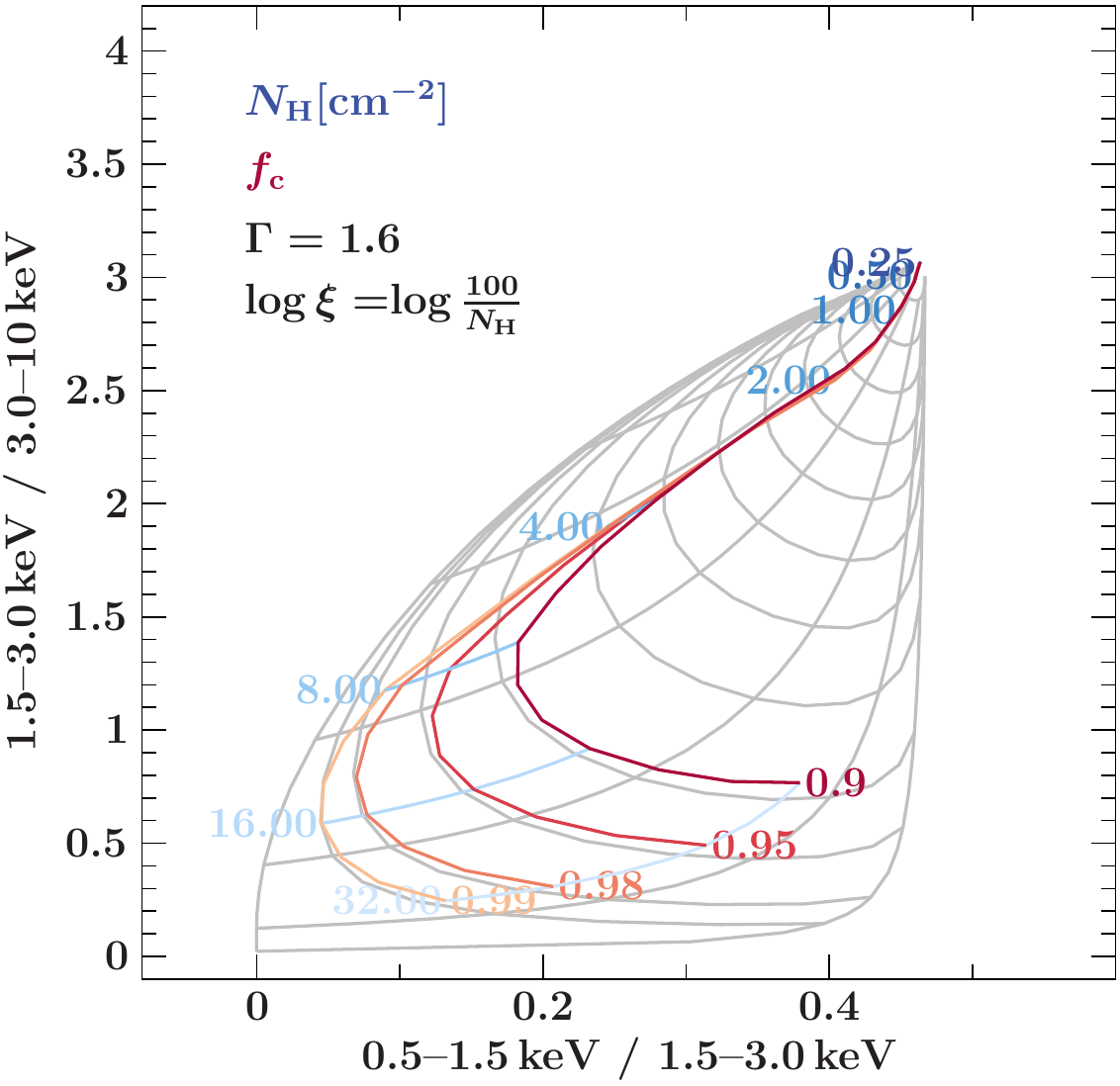}\hfill
    \includegraphics[width=0.31\textwidth]{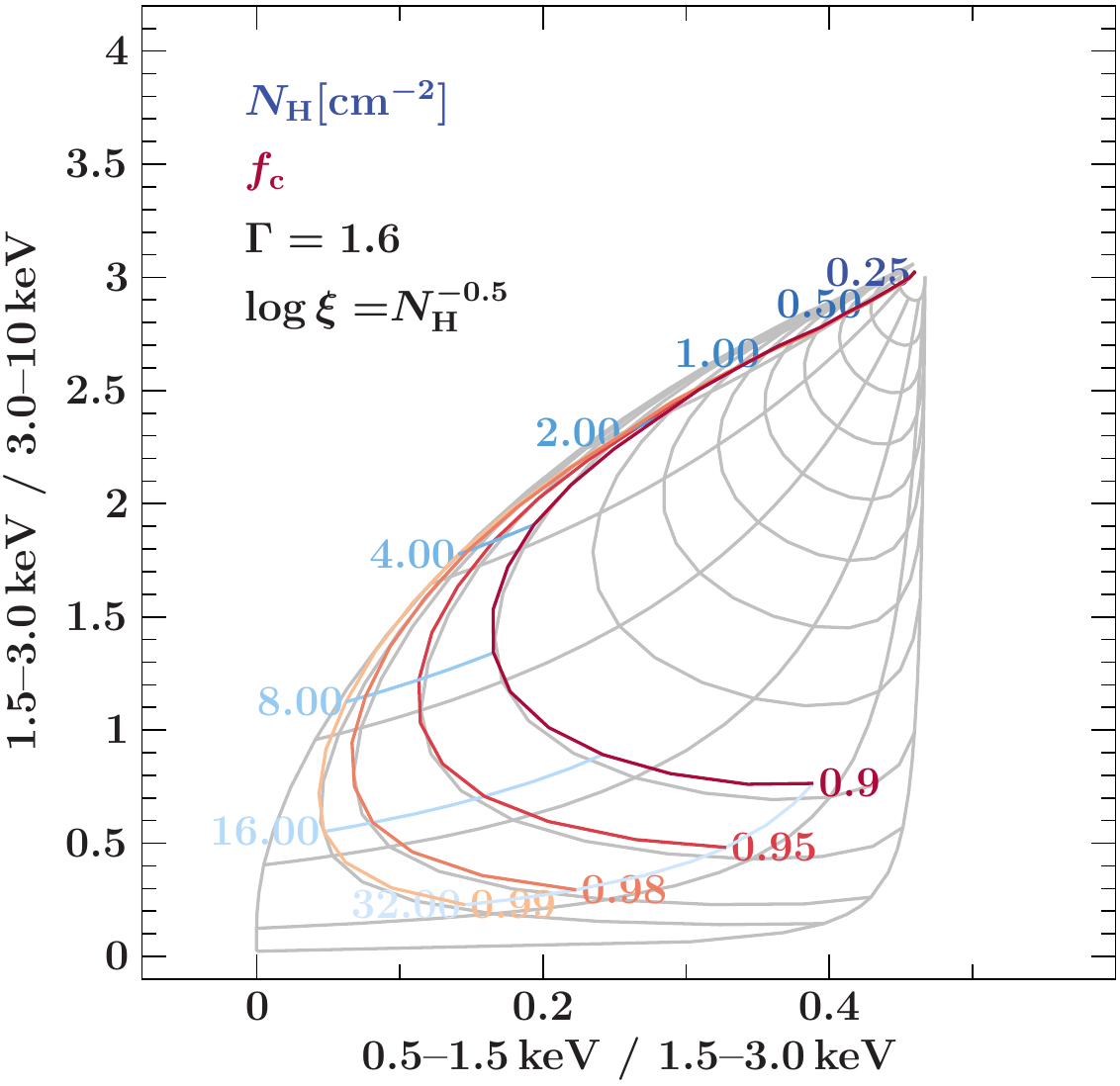}\hfill
        \includegraphics[width=0.31\textwidth]{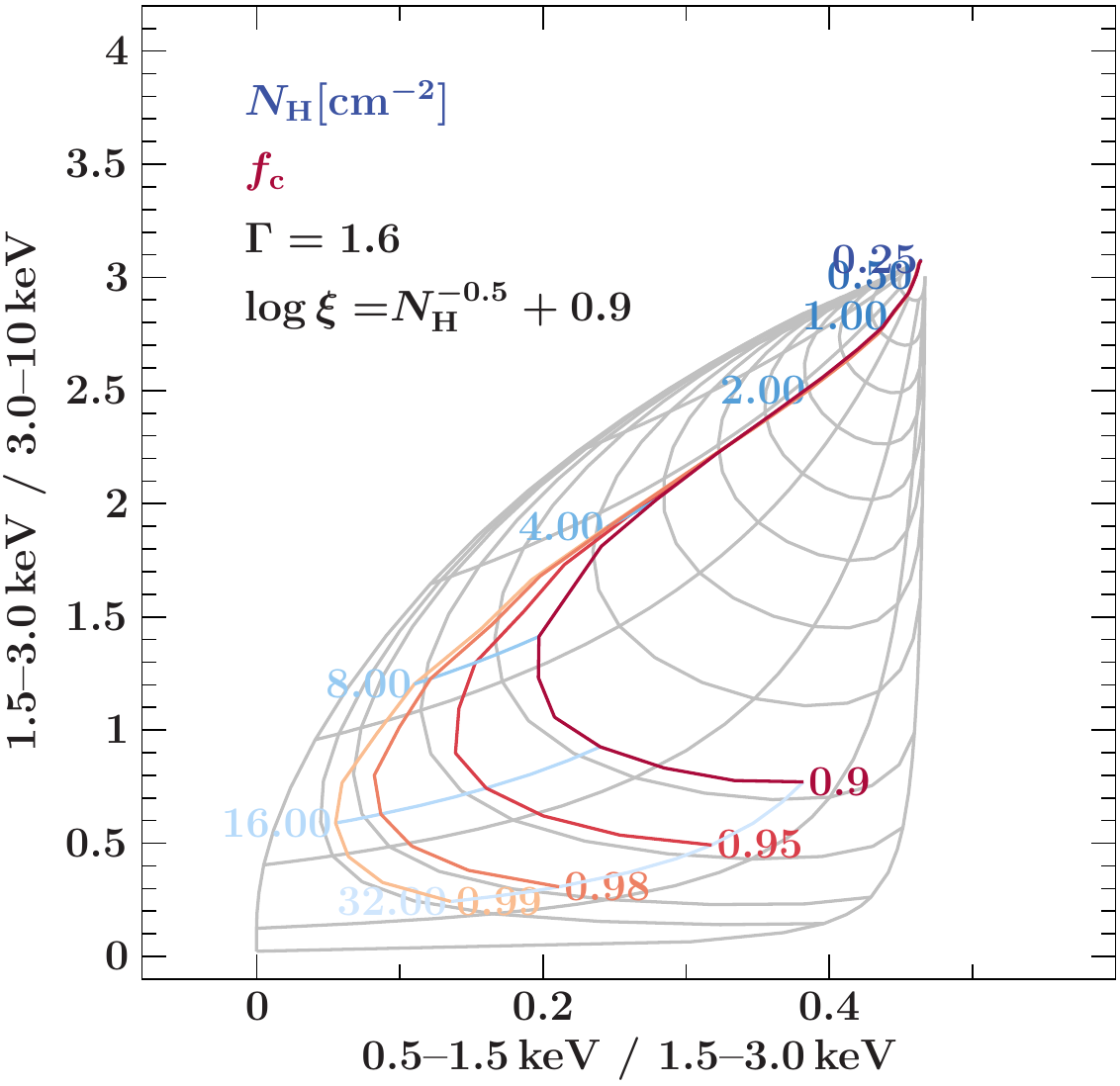}
    \caption{Imhomogeneos warm absorber grids for HETG-MEG with typical Cyg X-1 continuum parameters for different 
    dependencies of the ionization parameter $\log \xi$ on the equivalent hydrogen column density of the absorber, $N_\mathrm{H}$, as expressed in units of $10^{22}\,\mathrm{cm}^{-2}$. In all cases, a grid for the quasi-neutral absorption with \texttt{warmabs} ($\log xi = -4$) is shown in gray in the background to guide the eye. \textsl{Left:} case 1, $\log \xi = \log \frac{100}{N_\mathrm{H}}$. \textsl{Middle:} case 2, $\log \xi = N_\mathrm{H}^{-0.5}$. \textsl{Right:} case 3, $\log \xi = N_\mathrm{H}^{-0.5} + 0.9$.}
    \label{fig:xiofnh}
\end{figure*}

We show the resulting tracks for all three cases in Fig~\ref{fig:xiofnh}, where we also  compare them to quasi-neutral absorption as calculated for $\log \xi = -4$, i.e., the lowest absorption accessible for the \texttt{warmabs} models. It can be easily seen that for cases 1 and 3 the resulting tracks show a stronger curvature, as expected. This is not the case for case 2 that closely follows the tracks of the quasi-neutral case. This can be easily understood using Fig.~\ref{fig:logxiofnh}: case 2 only covers the interesting range of $\log \xi$, where strong changes in tracks are expected (cf. Fig.~\ref{fig:warmgrid}) for small values of $N_\mathrm{H}$, below $\sim10^{22}\,\mathrm{cm}^{-2}$, i.e., the absorbing material is not ionized enough for ionization effects to become visible on color-color tracks.

\subsubsection{Comparison to observed color-color tracks for Cyg X-1}

\begin{figure}
\includegraphics[width=\columnwidth]{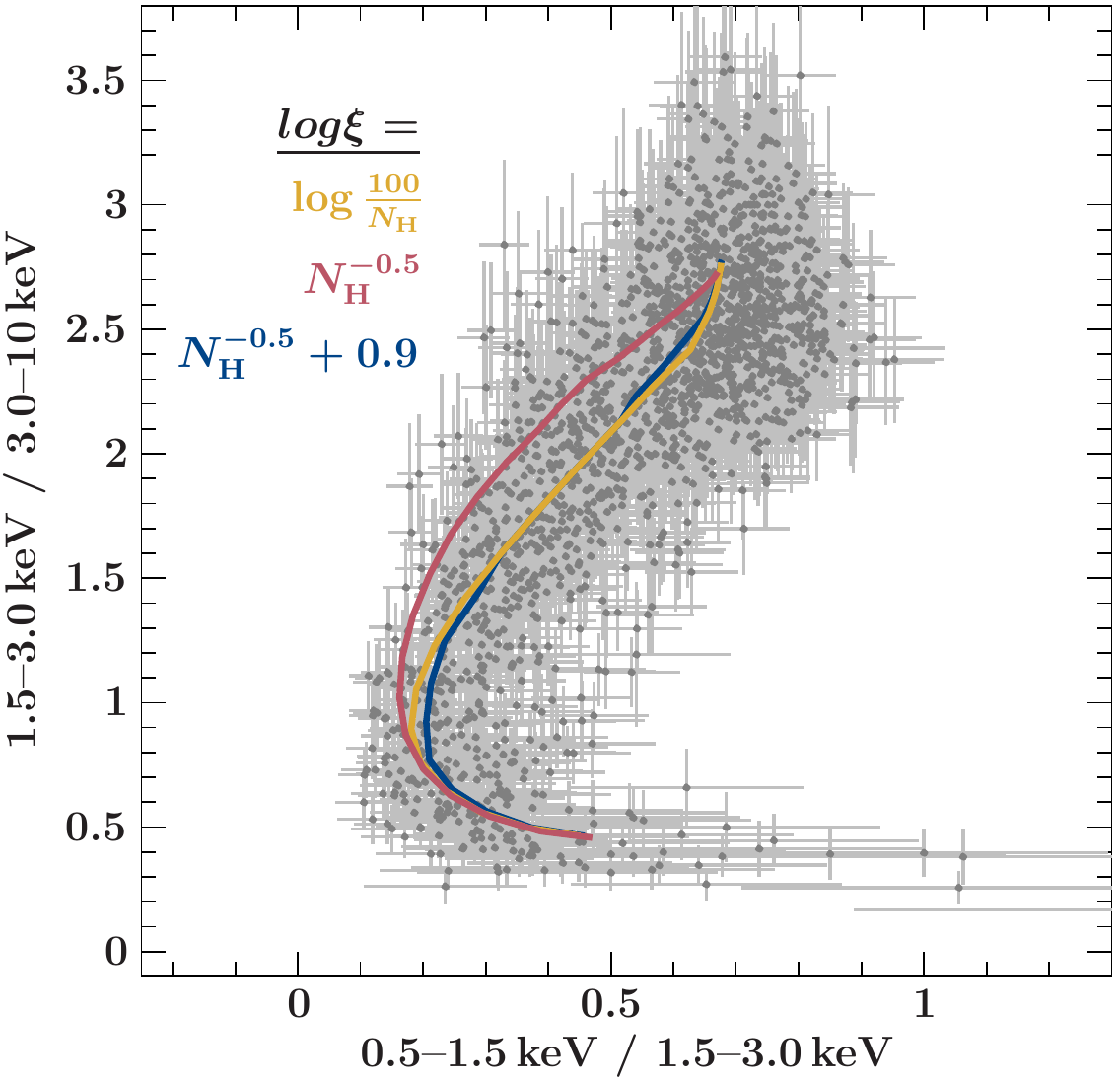}
\caption{Comparison of the color-color diagrams of the Chandra observation ObsID 3814 using the MEG M1 arm with theoretical tracks with different dependencies of the ionization parameter $\log \xi$ on the equivalent hydrogen column density, $N_\mathrm{N}$ expressed in units of $10^{22}\,\mathrm{cm}^{-2}$.}\label{fig:final}
\end{figure}

We compare the tracks for cases 1, 2, and 3 to the \textsl{Chandra} observations in Fig.~\ref{fig:final}, using $\Gamma = 1.6$ and $f_\mathrm{c} = 0.95$ as previously. Cases 1 and 3 describe the data well. Case 2 fails to describe the curvature of the data, as expected given that is does not cover the interesting range of the ionization parameter. We visually compare the tracks for case 1 and 3 with those obtained for constant ionization (Fig.~\ref{fig:chandra}, right panel): the curvature of the data is reproduced better by the two theoretical tracks with variable ionization.

This result is in agreement with our expectations of structured wind clumps with more ionized shells and less ionized cores. In such a set up,  the lesser absorption episodes are then caused by either smaller clumps in our line of sight or an outer part of a larger clump crossing the line of sight, %we thus see less ionized material. 
% MAN the following, right? -- VG: yes! Sorry, mine was obviously wrong way around.
and thus we see more highly ionized material.
The deepest absorption events are then caused by looking through the middle of the largest clumps,
% MAN correct? -- VG: yes
which are likely self-shielded and therefore exhibit lower ionization.

\section{Summary and Outlook}
\label{sec:outlook}

% NH: so, remember that deep-dip spectrum from either Hirsch 2019 or Hanke 2011, where fluorescence lines appear at low energies? Hirsch has a discussion about that one, where the idea was basically that we are more or less fully covered, i.e., absorbed, such that now we see the fluoresence emission from like the larger environment. With the assumption that we can't really absorb any further, I think we tried to use that and clump size assumptions to determine clump distance? Or Halo size? 
% Anyways, the fluorescence is additional flux. warmabs only includes absorption. So the natural next question is: how does that fluorescence fuck with your CC diagram? 
% I guess the emission would mostly add to the A band (<~1-1.5 keV, i.e., O and Ne). So probably that shifts tracks to the right in the diagram. Don't know about shape, though. 

We have shown that the typical curved tracks that wind-accretion HMXBs describe on color-color diagrams during absorption events (dips) can be explained in terms of a partial covering model with variable absorption. In particular, the tracks are due to changes in the effective absorbing column density and, for reasonable assumption of a power-law-like continuum shape, not due to variability of other model parameters.

Taking into account the ionization of the absorbing wind material is crucial for a realistic modelling of the color-color tracks. To do so we used the \texttt{warmabs} family of models, with both constant ionization parameter and an ionization parameter that depends on the equivalent absorbing column density. The second set up can be explained as dense wind clumps embedded in tenuous, hot inter-clump material and irradiated by X-rays from the compact object.

In particular, we compare with Chandra/HETG observation of the HMXB Cyg X-1 and show that both the neutral absorber cannot explain the curvature of the tracks. A homogeneously ionized absorber with a mild ionization of $\log \xi \approx 1.6$ fares better. The best description is achieved if the ionization of the absorbing material is a function of the absorbing column density, specifically if the ionization decreases with increasing column.
% MAN This is where you should be prepared for the referee to say "Hey, could you MCMC your results???" -- VG: moved it into the outlook to make it more clear that I don't want to do it NOW

The approach presented here opens several avenues for further work. Color-color diagrams could be used to test whether the material causing the increase of absorption in the cases where detailed high resolution spectroscopic analysis are not available, e.g., due to the use of CCD based instruments or due to low count rates either because of intrinsic source faintness or due to the short duration of dipping events.
%is ionized without having to undertake detailed high-resolution spectroscopic analyses to trace the individual elements, 
% NH: this phrasing is just a little ambiguous in whether you are saying use high-res obs for the CC diagram anyway, or high-res isn't available so do CC diagrams with the low-res instruments. The natural follow-up question to that would be: what happens to your traces if you do this with CCD resolution instead? Any impact on conclusions you can draw from the CC digrams if they have low spectral resolution? 
% I supposed, since the colors are technically spectra added up over extra low resolution (3 bands), it shouldn't matter too much (hough there is some low-res instrument induced smearing across the edges between energy bands). And the CC in Nowak 2008 from Suzaku looks comparable. 
%as high quality high resolution observations during absorption events are not accessible in many fainter sources or in sources where absorption events are rare. 
Further, \citet{El_Mellah_2020a} have recently presented the theoretical framework for a formalism that, based on simulations of different line of  sights through simplified clumpy winds, connects short-term absorption variability with clump size and mass. Given observations of sufficient length at a given orbital phase, color-color tracks could be used to obtain absorption light curves with enough time resolution to constrain absorption variability and thus to test their clumpy wind models. %Furthermore, future instruments, from high throughput missions such as \textsl{eXTP} to sensitive micro-calorimeters onboard \textsl{XRISM} and \textsl{Athena}, will make currently too faint HMXBs accessible for absorption studies such as presented here.

%for observations of sufficient length at a given orbital that could be obtained in the future.
 
% Higher sensitivity of future instruments such as XRISM and Athena, means more sources where we can actually see and disentangle this behavior. Athena: bright sources in a few 100s (Lomaeva), but not faint sources, will still need color-color.

\begin{acknowledgements} VG is supported through the Margarete von Wrangell fellowship by the ESF and the Ministry of Science, Research and the Arts Baden-W\"urttemberg. Work at LLNL was performed under the auspices of the U.S. Department of Energy under Contract No.\ DE-AC52-07NA27344 and supported by NASA grants to LLNL. 
MAN was supported by NASA Grant NNX12AE37G in earlier stages of this work.
This research has made
  use of NASA's
  Astrophysics Data System Bibliographic Service (ADS) and of ISIS functions (\texttt{isisscripts})\footnote{\url{http://www.sternwarte.uni-erlangen.de/isis/}}
  provided by ECAP/Remeis observatory and MIT. We in particular thank Mirjam Oertel, who developed the first versions of some of the scripts calculating the color-color tracks for arbitrary input functions, and John
  E.  Davis for the development of the
  \texttt{slxfig}\footnote{\url{http://www.jedsoft.org/fun/slxfig/}}
  module used to prepare the figures in this work. Some of the
  color schemes used were based on Paul Tol's color-blindness-friendly palettes and
  templates\footnote{\url{https://personal.sron.nl/~pault/}}. 
\end{acknowledgements}

\bibliographystyle{aa}
\bibliography{aa_abbrv,mnemonic,references}

\begin{thebibliography}{53}
\expandafter\ifx\csname natexlab\endcsname\relax\def\natexlab#1{#1}\fi

\bibitem[{{Arnaud}(1996)}]{Arnaud_1996a}
{Arnaud}, K.~A. 1996, Astronomical Society of the Pacific Conference Series,
  Vol. 101, {XSPEC: The First Ten Years}, ed. G.~H. {Jacoby} \& J.~{Barnes}, 17

\bibitem[{{Ba{\l}uci{\'n}ska-Church} {et~al.}(2000){Ba{\l}uci{\'n}ska-Church},
  {Church}, {Charles}, {Nagase}, {LaSala}, \&
  {Barnard}}]{Balucinska-Church_2000a}
{Ba{\l}uci{\'n}ska-Church}, M., {Church}, M.~J., {Charles}, P.~A., {et~al.}
  2000, \mnras, 311, 861

\bibitem[{{Basak} {et~al.}(2017){Basak}, {Zdziarski}, {Parker}, \&
  {Islam}}]{Basak_2017a}
{Basak}, R., {Zdziarski}, A.~A., {Parker}, M., \& {Islam}, N. 2017, \mnras,
  472, 4220

\bibitem[{{Bautista} \& {Kallman}(2001)}]{Bautista_2001a}
{Bautista}, M.~A. \& {Kallman}, T.~R. 2001, \apjs, 134, 139

\bibitem[{{Blondin} {et~al.}(1990){Blondin}, {Kallman}, {Fryxell}, \&
  {Taam}}]{Blondin_1990a}
{Blondin}, J.~M., {Kallman}, T.~R., {Fryxell}, B.~A., \& {Taam}, R.~E. 1990,
  \apj, 356, 591

\bibitem[{{Boroson} {et~al.}(2003){Boroson}, {Vrtilek}, {Kallman}, \&
  {Corcoran}}]{Boroson_2003a}
{Boroson}, B., {Vrtilek}, S.~D., {Kallman}, T., \& {Corcoran}, M. 2003, \apj,
  592, 516

\bibitem[{{Bozzo} {et~al.}(2011){Bozzo}, {Giunta}, {Cusumano}, {Ferrigno},
  {Walter}, {Campana}, {Falanga}, {Israel}, \& {Stella}}]{Bozzo_2011a}
{Bozzo}, E., {Giunta}, A., {Cusumano}, G., {et~al.} 2011, \aap, 531, A130

\bibitem[{{Carpano} {et~al.}(2005){Carpano}, {Wilms}, {Schirmer}, \&
  {Kendziorra}}]{Carpano_2005a}
{Carpano}, S., {Wilms}, J., {Schirmer}, M., \& {Kendziorra}, E. 2005, \aap,
  443, 103

\bibitem[{{Cohen} {et~al.}(2011){Cohen}, {Gagn{\'e}}, {Leutenegger},
  {MacArthur}, {Wollman}, {Sundqvist}, {Fullerton}, \& {Owocki}}]{Cohen_2011a}
{Cohen}, D.~H., {Gagn{\'e}}, M., {Leutenegger}, M.~A., {et~al.} 2011, \mnras,
  415, 3354

\bibitem[{{El Mellah} {et~al.}(2020){El Mellah}, {Grinberg}, {Sundqvist},
  {Driessen}, \& {Leutenegger}}]{El_Mellah_2020a}
{El Mellah}, I., {Grinberg}, V., {Sundqvist}, J.~O., {Driessen}, F.~A., \&
  {Leutenegger}, M.~A. 2020, arXiv e-prints, arXiv:2006.16216

\bibitem[{{El Mellah} {et~al.}(2018){El Mellah}, {Sundqvist}, \&
  {Keppens}}]{El_Mellah_2018a}
{El Mellah}, I., {Sundqvist}, J.~O., \& {Keppens}, R. 2018, \mnras, 475, 3240

\bibitem[{{Feldmeier}(1995)}]{Feldmeier_1995a}
{Feldmeier}, A. 1995, \aap, 299, 523

\bibitem[{{Ferrigno} {et~al.}(2020){Ferrigno}, {Bozzo}, \&
  {Romano}}]{Ferrigno_2020a}
{Ferrigno}, C., {Bozzo}, E., \& {Romano}, P. 2020, arXiv e-prints,
  arXiv:2008.04657

\bibitem[{{Fornasini} {et~al.}(2017){Fornasini}, {Tomsick}, {Bachetti},
  {Krivonos}, {F{\"u}rst}, {Natalucci}, {Pottschmidt}, \&
  {Wilms}}]{Fornasini_2017a}
{Fornasini}, F.~M., {Tomsick}, J.~A., {Bachetti}, M., {et~al.} 2017, \apj, 841,
  35

\bibitem[{{Fullerton} {et~al.}(2006){Fullerton}, {Massa}, \&
  {Prinja}}]{Fullerton_2006a}
{Fullerton}, A.~W., {Massa}, D.~L., \& {Prinja}, R.~K. 2006, \apj, 637, 1025

\bibitem[{{F{\"u}rst} {et~al.}(2010){F{\"u}rst}, {Kreykenbohm}, {Pottschmidt},
  {Wilms}, {Hanke}, {Rothschild}, {Kretschmar}, {Schulz}, {Huenemoerder},
  {Klochkov}, \& {Staubert}}]{Fuerst_2010a}
{F{\"u}rst}, F., {Kreykenbohm}, I., {Pottschmidt}, K., {et~al.} 2010, \aap,
  519, A37

\bibitem[{{F{\"u}rst} {et~al.}(2014){F{\"u}rst}, {Pottschmidt}, {Wilms},
  {Tomsick}, {Bachetti}, {Boggs}, {Christensen}, {Craig}, {Grefenstette},
  {Hailey}, {Harrison}, {Madsen}, {Miller}, {Stern}, {Walton}, \&
  {Zhang}}]{Fuerst_2014a}
{F{\"u}rst}, F., {Pottschmidt}, K., {Wilms}, J., {et~al.} 2014, \apj, 780, 133

\bibitem[{{Garc{\'\i}a} {et~al.}(2018){Garc{\'\i}a}, {Fogantini}, {Chaty}, \&
  {Combi}}]{Garcia_2018a}
{Garc{\'\i}a}, F., {Fogantini}, F.~A., {Chaty}, S., \& {Combi}, J.~A. 2018,
  \aap, 618, A61

\bibitem[{{Grinberg} {et~al.}(2017){Grinberg}, {Hell}, {El Mellah}, {Neilsen},
  {Sander}, {Leutenegger}, {F{\"u}rst}, {Huenemoerder}, {Kretschmar},
  {K{\"u}hnel}, {Mart{\'{\i}}nez-N{\'u}{\~n}ez}, {Niu}, {Pottschmidt},
  {Schulz}, {Wilms}, \& {Nowak}}]{Grinberg_2017a}
{Grinberg}, V., {Hell}, N., {El Mellah}, I., {et~al.} 2017, \aap, 608, A143

\bibitem[{{Grinberg} {et~al.}(2013){Grinberg}, {Hell}, {Pottschmidt},
  {B{\"o}ck}, {Nowak}, {Rodriguez}, {Bodaghee}, {Cadolle Bel}, {Case}, {Hanke},
  {K{\"u}hnel}, {Markoff}, {Pooley}, {Rothschild}, {Tomsick}, {Wilson-Hodge},
  \& {Wilms}}]{Grinberg_2013a}
{Grinberg}, V., {Hell}, N., {Pottschmidt}, K., {et~al.} 2013, \aap, 554, A88

\bibitem[{{Grinberg} {et~al.}(2015){Grinberg}, {Leutenegger}, {Hell},
  {Pottschmidt}, {B{\"o}ck}, {Garc{\'{\i}}a}, {Hanke}, {Nowak}, {Sundqvist},
  {Townsend}, \& {Wilms}}]{Grinberg_2015a}
{Grinberg}, V., {Leutenegger}, M.~A., {Hell}, N., {et~al.} 2015, \aap, 576,
  A117

\bibitem[{{Haberl} {et~al.}(1989){Haberl}, {White}, \&
  {Kallman}}]{Haberl_1989a}
{Haberl}, F., {White}, N.~E., \& {Kallman}, T.~R. 1989, \apj, 343, 409

\bibitem[{{Hanke}(2011)}]{Hanke_2011_PhD}
{Hanke}, M. 2011, PhD thesis, Dr.~Karl Remeis-Sternwarte, Astronomisches
  Institut der Universit{\"a}t Erlangen-N{\"u}rnberg, Sternwartstr.~7, 96049
  Bamberg, Germany

\bibitem[{{Hanke} {et~al.}(2009){Hanke}, {Wilms}, {Nowak}, {Pottschmidt},
  {Schulz}, \& {Lee}}]{Hanke_2009a}
{Hanke}, M., {Wilms}, J., {Nowak}, M.~A., {et~al.} 2009, \apj, 690, 330

\bibitem[{{Hanke} {et~al.}(2008){Hanke}, {Wilms}, {Nowak}, {Schulz},
  {Pottschmidt}, {Lee}, \& {Boeck}}]{Hanke_2008a}
{Hanke}, M., {Wilms}, J., {Nowak}, M.~A., {et~al.} 2008, in Microquasars and
  Beyond

\bibitem[{{Hemphill} {et~al.}(2014){Hemphill}, {Rothschild}, {Markowitz},
  {F{\"u}rst}, {Pottschmidt}, \& {Wilms}}]{Hemphill_2014a}
{Hemphill}, P.~B., {Rothschild}, R.~E., {Markowitz}, A., {et~al.} 2014, \apj,
  792, 14

\bibitem[{{Hickox} {et~al.}(2004){Hickox}, {Narayan}, \&
  {Kallman}}]{Hickox_2004a}
{Hickox}, R.~C., {Narayan}, R., \& {Kallman}, T.~R. 2004, \apj, 614, 881

\bibitem[{{Hirsch} {et~al.}(2019){Hirsch}, {Hell}, {Grinberg}, {Ballhausen},
  {Nowak}, {Pottschmidt}, {Schulz}, {Dauser}, {Hanke}, {Kallman}, {Brown}, \&
  {Wilms}}]{Hirsch_2019a}
{Hirsch}, M., {Hell}, N., {Grinberg}, V., {et~al.} 2019, \aap, 626, A64

\bibitem[{{Houck}(2002)}]{Houck_2002}
{Houck}, J.~C. 2002, in High Resolution X-ray Spectroscopy with XMM-Newton and
  Chandra, ed. {G.~Branduardi-Raymont}

\bibitem[{{Houck} \& {Denicola}(2000)}]{Houck_Denicola_2000a}
{Houck}, J.~C. \& {Denicola}, L.~A. 2000, in Astronomical Society of the
  Pacific Conference Series, Vol. 216, Astronomical Data Analysis Software and
  Systems IX, ed. {N.~Manset, C.~Veillet, \& D.~Crabtree}, 591

\bibitem[{{Kallman} \& {Bautista}(2001)}]{Kallman_2001a}
{Kallman}, T. \& {Bautista}, M. 2001, \apjs, 133, 221

\bibitem[{{Kallman} {et~al.}(2009){Kallman}, {Bautista}, {Goriely}, {Mendoza},
  {Miller}, {Palmeri}, {Quinet}, \& {Raymond}}]{Kallman_2009a}
{Kallman}, T.~R., {Bautista}, M.~A., {Goriely}, S., {et~al.} 2009, \apj, 701,
  865

\bibitem[{{Li} \& {Clark}(1974)}]{Li_1974a}
{Li}, F.~K. \& {Clark}, G.~W. 1974, \apjl, 191, L27

\bibitem[{{Lomaeva} {et~al.}(2020){Lomaeva}, {Grinberg}, {Guainazzi}, {Hell},
  {Bianchi}, {K{\"u}hnel}, {F{\"u}rst}, {Kretschmar},
  {Mart{\'\i}nez-Chicharro}, {Mart{\'\i}nez-N{\'u}{\~n}ez}, \&
  {Torrej{\'o}n}}]{Lomaeva_2020a}
{Lomaeva}, M., {Grinberg}, V., {Guainazzi}, M., {et~al.} 2020, arXiv e-prints,
  arXiv:2007.07260

\bibitem[{{Mart{\'{\i}}nez-N{\'u}{\~n}ez}
  {et~al.}(2017){Mart{\'{\i}}nez-N{\'u}{\~n}ez}, {Kretschmar}, {Bozzo},
  {Oskinova}, {Puls}, {Sidoli}, {Sundqvist}, {Blay}, {Falanga}, {F{\"u}rst},
  {G{\'{\i}}menez-Garc{\'{\i}}a}, {Kreykenbohm}, {K{\"u}hnel}, {Sander},
  {Torrej{\'o}n}, \& {Wilms}}]{Martinez-Nunez_2017a}
{Mart{\'{\i}}nez-N{\'u}{\~n}ez}, S., {Kretschmar}, P., {Bozzo}, E., {et~al.}
  2017, \ssr, 212, 59

\bibitem[{{Mart{\'{\i}}nez-N{\'u}{\~n}ez}
  {et~al.}(2014){Mart{\'{\i}}nez-N{\'u}{\~n}ez}, {Torrej{\'o}n}, {K{\"u}hnel},
  {Kretschmar}, {Stuhlinger}, {Rodes-Roca}, {F{\"u}rst}, {Kreykenbohm},
  {Martin-Carrillo}, {Pollock}, \& {Wilms}}]{Martinez-Nunez_2014a}
{Mart{\'{\i}}nez-N{\'u}{\~n}ez}, S., {Torrej{\'o}n}, J.~M., {K{\"u}hnel}, M.,
  {et~al.} 2014, \aap, 563, A70

\bibitem[{{Mi{\v s}kovi{\v c}ov{\'a}} {et~al.}(2016){Mi{\v s}kovi{\v
  c}ov{\'a}}, {Hell}, {Hanke}, {Nowak}, {Pottschmidt}, {Schulz}, {Grinberg},
  {Duro}, {Madej}, {Lohfink}, {Rodriguez}, {Cadolle Bel}, {Bodaghee},
  {Tomsick}, {Lee}, {Brown}, \& {Wilms}}]{Miskovicova_2016a}
{Mi{\v s}kovi{\v c}ov{\'a}}, I., {Hell}, N., {Hanke}, M., {et~al.} 2016, \aap,
  590, A114

\bibitem[{{Naik} {et~al.}(2011){Naik}, {Paul}, \& {Ali}}]{Naik_2011a}
{Naik}, S., {Paul}, B., \& {Ali}, Z. 2011, \apj, 737, 79

\bibitem[{{Noble} \& {Nowak}(2008)}]{Noble_Nowak_2008a}
{Noble}, M.~S. \& {Nowak}, M.~A. 2008, \pasp, 120, 821

\bibitem[{{Nowak} {et~al.}(2011){Nowak}, {Hanke}, {Trowbridge}, {Markoff},
  {Wilms}, {Pottschmidt}, {Coppi}, {Maitra}, {Davis}, \&
  {Tramper}}]{Nowak_2011a}
{Nowak}, M.~A., {Hanke}, M., {Trowbridge}, S.~N., {et~al.} 2011, \apj, 728, 13

\bibitem[{{Odaka} {et~al.}(2013){Odaka}, {Khangulyan}, {Tanaka}, {Watanabe},
  {Takahashi}, \& {Makishima}}]{Odaka_2013a}
{Odaka}, H., {Khangulyan}, D., {Tanaka}, Y.~T., {et~al.} 2013, \apj, 767, 70

\bibitem[{{Oskinova} {et~al.}(2012){Oskinova}, {Feldmeier}, \&
  {Kretschmar}}]{Oskinova_2012a}
{Oskinova}, L.~M., {Feldmeier}, A., \& {Kretschmar}, P. 2012, \mnras, 421, 2820

\bibitem[{{Owocki} \& {Rybicki}(1984)}]{Owocki_1984a}
{Owocki}, S.~P. \& {Rybicki}, G.~B. 1984, \apj, 284, 337

\bibitem[{{Sander} {et~al.}(2018){Sander}, {F{\"u}rst}, {Kretschmar},
  {Oskinova}, {Todt}, {Hainich}, {Shenar}, \& {Hamann}}]{Sander_2018a}
{Sander}, A.~A.~C., {F{\"u}rst}, F., {Kretschmar}, P., {et~al.} 2018, \aap,
  610, A60

\bibitem[{{Skipper} {et~al.}(2013){Skipper}, {McHardy}, \&
  {Maccarone}}]{Skipper_2013a}
{Skipper}, C.~J., {McHardy}, I.~M., \& {Maccarone}, T.~J. 2013, \mnras, 434,
  574

\bibitem[{{Sundqvist} \& {Owocki}(2013)}]{Sundqvist_2013a}
{Sundqvist}, J.~O. \& {Owocki}, S.~P. 2013, \mnras, 428, 1837

\bibitem[{{Sundqvist} {et~al.}(2018){Sundqvist}, {Owocki}, \&
  {Puls}}]{Sundqvist_2018a}
{Sundqvist}, J.~O., {Owocki}, S.~P., \& {Puls}, J. 2018, \aap, 611, A17

\bibitem[{{Tarter} {et~al.}(1969){Tarter}, {Tucker}, \&
  {Salpeter}}]{Tarter_1969a}
{Tarter}, C.~B., {Tucker}, W.~H., \& {Salpeter}, E.~E. 1969, \apj, 156, 943

\bibitem[{{Torrej{\'o}n} {et~al.}(2015){Torrej{\'o}n}, {Schulz}, {Nowak},
  {Oskinova}, {Rodes-Roca}, {Shenar}, \& {Wilms}}]{Torrejon_2015a}
{Torrej{\'o}n}, J.~M., {Schulz}, N.~S., {Nowak}, M.~A., {et~al.} 2015, \apj,
  810, 102

\bibitem[{{Verner} {et~al.}(1996){Verner}, {Ferland}, {Korista}, \&
  {Yakovlev}}]{Verner_1996a}
{Verner}, D.~A., {Ferland}, G.~J., {Korista}, K.~T., \& {Yakovlev}, D.~G. 1996,
  \apj, 465, 487

\bibitem[{{Wilms} {et~al.}(2000){Wilms}, {Allen}, \& {McCray}}]{Wilms_2000a}
{Wilms}, J., {Allen}, A., \& {McCray}, R. 2000, \apj, 542, 914

\bibitem[{{Xu} {et~al.}(1986){Xu}, {McCray}, \& {Kelley}}]{Xu_1986a}
{Xu}, Y., {McCray}, R., \& {Kelley}, R. 1986, \nat, 319, 652

\bibitem[{{Yamada} {et~al.}(2013){Yamada}, {Torii}, {Mineshige}, {Ueda},
  {Kubota}, {Gandhi}, {Done}, {Noda}, {Yoshikawa}, \&
  {Makishima}}]{Yamada_2013a}
{Yamada}, S., {Torii}, S., {Mineshige}, S., {et~al.} 2013, \apjl, 767, L35

\end{thebibliography}

\end{document}